\documentclass[aps,pra,twocolumn,groupedaddress,floatfix,superscriptaddress,longbibliography]{revtex4-1}

\usepackage{amsmath}
\usepackage{amssymb}
\usepackage{graphicx}
\usepackage{float}
\usepackage[english]{babel}
\usepackage{dsfont}
\usepackage{epstopdf}
\usepackage{soul}
\usepackage[normalem]{ulem}
\usepackage{color}
\usepackage{braket}
\usepackage{upgreek}
\usepackage{bm}
\usepackage[colorlinks=true,
            linkcolor=magenta,
            urlcolor=blue,
            citecolor=blue]{hyperref}

\usepackage[normalem]{ulem}

\def\be{\begin{equation}}
\def\ee{\end{equation}} 
\def\bea{\begin{eqnarray}}
\def\eea{\end{eqnarray}} 
\def\ba{\begin{array}} 
\def\ea{\end{array}}

\emergencystretch=\maxdimen
\begin{document}
\title{Quantized valley Hall response from local bulk density variations}

\author{Maxime Jamotte}
\email{mjamotte@ulb.be}
\affiliation{Center for Nonlinear Phenomena and Complex Systems, Universit\'e Libre de Bruxelles, CP 231, Campus Plaine, B-1050 Brussels, Belgium}
\author{Lucila Peralta Gavensky}
\email{lucila.peralta.gavensky@ulb.be}
\affiliation{Center for Nonlinear Phenomena and Complex Systems, Universit\'e Libre de Bruxelles, CP 231, Campus Plaine, B-1050 Brussels, Belgium}
\author{\mbox{Cristiane Morais Smith}}
\affiliation{Institute for Theoretical Physics, Utrecht University, Princetonplein 5, 3584CC Utrecht, The Netherlands}
\author{Marco Di Liberto}
\email{marco.diliberto@unipd.it}
\affiliation{Dipartimento di Fisica e Astronomia ``G. Galilei" \& Padua Quantum Technologies
Research Center, Università degli Studi di Padova, I-35131, Padova, Italy}
\affiliation{INFN Istituto Nazionale di Fisica Nucleare, Sezione di Padova, I-35131, Padova, Italy}
\author{Nathan Goldman}
\email{ngoldman@ulb.be}
\affiliation{Center for Nonlinear Phenomena and Complex Systems, Universit\'e Libre de Bruxelles, CP 231, Campus Plaine, B-1050 Brussels, Belgium}

\maketitle

\section*{Abstract}
\textbf{The application of a mechanical strain to a 2D material can create pseudo-magnetic fields and lead to a quantized valley Hall effect. However, measuring valley-resolved effects remains a challenging task due to their inherent fragility and dependence on the sample's proper design. Additionally, non-local transport probes based on multiterminal devices have often proven to be inadequate in yielding conclusive evidence of the valley Hall signal. Here, we introduce an alternative way of detecting the quantized valley Hall effect, which entirely relies on local density measurements, performed deep in the bulk of the sample. The resulting quantized signal is a genuine Fermi sea response, independent of the edge physics, and reflects the underlying valley Hall effect through the Widom-St\v{r}eda formula. Specifically, our approach is based on measuring the variation of the particle density, locally in the bulk, upon varying the strength of the applied strain. This approach to the quantized valley Hall effect is particularly well suited for experiments based on synthetic lattices, where the particle density (or integrated density of states) can be spatially resolved.}

\section*{Introduction}

    Two-dimensional honeycomb lattices, such as graphene or monolayers of transition metal dichalcogenides (TMDs), are considered to be promising candidates for valley based electronics owing to their particular bandstructure, which features two non-equivalent Dirac valleys at the corners of the Brillouin zone~\cite{Schaibley2016,Vitale2018}. Due to their large separation in momentum space, these constitute a novel discrete orbital degree of freedom for low-energy carriers, which could be used to store, filter and transport information~\cite{Rycerz2007} much in the same way the electronic spin is used for spintronics applications. When spatial inversion symmetry is explicitly~\cite{Xiao2007} or effectively~\cite{Marino2015} broken in these materials, while preserving time-reversal symmetry, finite but opposite Berry curvatures develop at each valley, endowing the system's carriers with an anomalous velocity. Indeed, in sufficiently clean samples with suppressed intervalley scattering, applying an in-plane electric field generates non-dissipative counterpropagating valley currents along its transverse direction without a net charge flow, giving rise to what is known as the valley Hall effect (VHE). Indirect measurements of this effect were reported in graphene on top of a hexagonal boron nitride substrate~\cite{Gorbachev2014}, biased graphene bilayers~\cite{Sui2015,Shimazaki2015} and TMDs~\cite{Mak2014}. Typical transport probes of the valley Hall conductivity are based on non-local resistance measurements in multiterminal Hall bar geometries, in a similar scheme as the one used to detect spin Hall currents~\cite{Abanin2011}. Despite early progress in interpreting the non-local resistance as a hallmark of the valley Hall effect, this analysis has been questioned by theoretical simulations~\cite{Kirczenow2015,Marmolejo-Tejada_2018,Aktor2021} and subsequent experiments~\cite{Zhu2017,Aharon-Steinberg2021}. In particular, the ongoing debate regarding the relative significance of edge versus bulk contributions in these early experiments, as well as the role of topological versus non-topological factors, continues to spark intense discussions in the scientific community~\cite{Torres_2021,Roche2022}. Other alternatives to these transport experiments require the use of spatially-resolved optical Kerr signals~\cite{Lee2016} or selectively exciting valley-polarized
carriers via circularly polarized light~\cite{Koppens2022}.

\begin{figure}[!t]
	\center
	\includegraphics[width=\columnwidth]{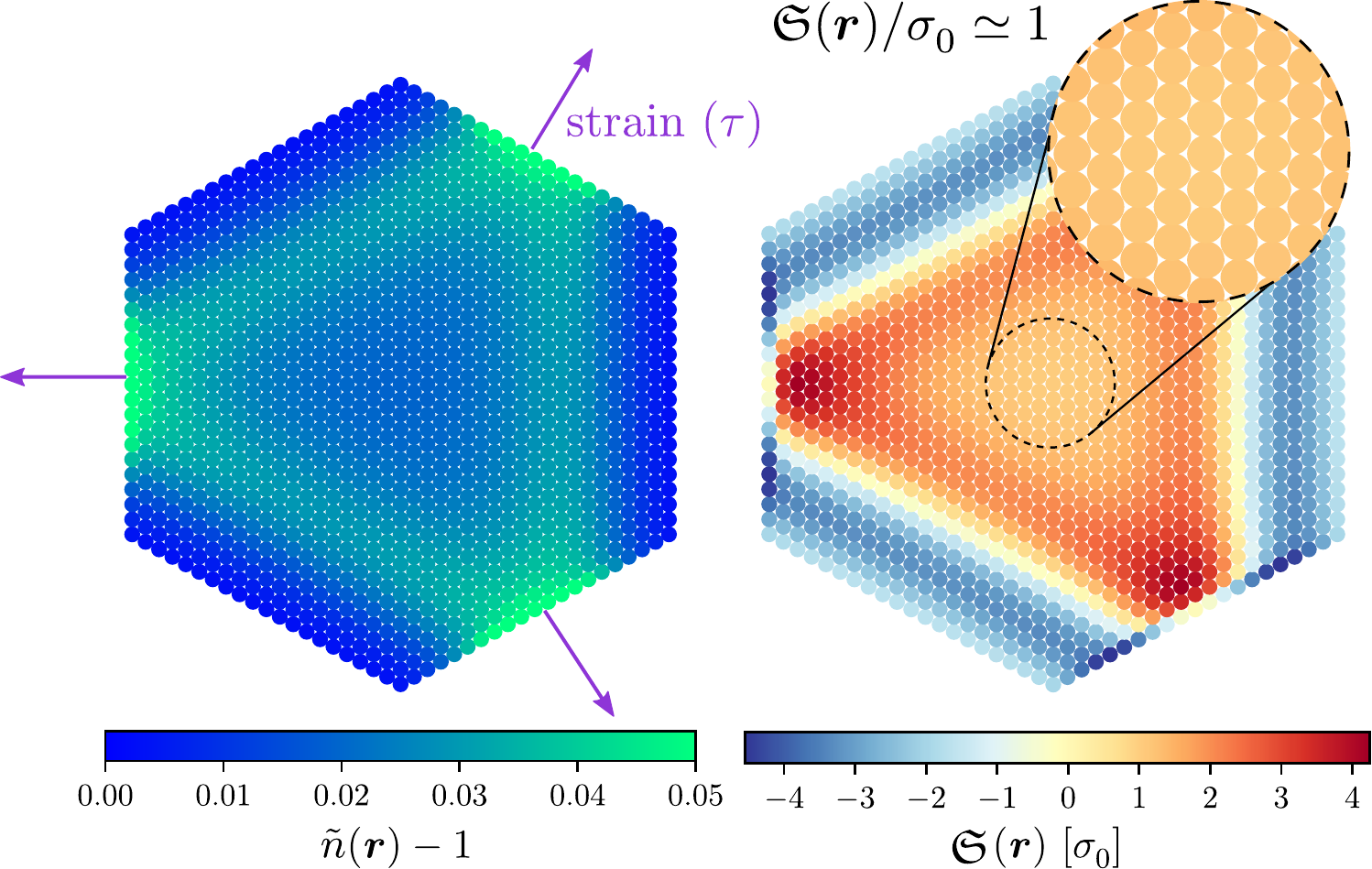}
	\caption{Measuring the variation of the particle density, upon varying the strength of the applied strain $\tau$, leads to a quantized bulk response, $\mathfrak{S}(\bm{r})$, reflecting the underlying valley Hall effect. Left panel: Number of particles $\tilde n(\bm{r})$ within a unit cell at position $\bm{r}$ when applying trigonal strain to the sample (arrows). Right panel: Local valley Hall marker  $\mathfrak{S}(\bm{r})$ [Eq.~\eqref{local_marker}] in units of the conductivity quantum $\sigma_0=e^2/h$, displaying a plateau at a quantized value $\mathfrak{S}(\boldsymbol{r})/\sigma_0^{} \in \mathbb{Z}$ deep in the bulk. In the present case, $\mathfrak S(\bm{r})/\sigma_0^{} \simeq 1$, as emphasized in the zoomed region. This local bulk response exists irrespective of the sample's edge termination. For more details on the system parameters used, see the Results section.}
	\label{Fig_1_story}
\end{figure}

Strained honeycomb lattices provide yet another platform where valley Hall related phenomena can take place. Interestingly, the coupling between the electronic degrees of freedom and the mechanical deformation in these samples can be well described by the emergence of strong artificial gauge fields, which drastically alter their low-energy properties~\cite{CastroNeto2009,Guinea2010a}. When properly engineering the strain, effective pseudomagnetic fields pointing in opposite directions develop at inequivalent Dirac cones, resulting in the formation of relativistic pseudo-Landau levels (pLLs) in the vicinity of the valleys. This characteristic energy spectrum has been successfully observed in a broad variety of devices: starting from strained graphene nanobubbles~\cite{Levy2010} and molecular graphene~\cite{Gardenier2020,Gomes2012}, all the way up to acoustic meta-materials~\cite{Abbaszadeh2017,Brendel2017,Wen2019}, polaritonic lattices~\cite{Jamadi2020}, honeycomb arrays of photonic waveguides~\cite{Rechtsman2013}, microwave resonators~\cite{Bellec2020} 
 and, more recently, photonic Fock-state lattices~\cite{Deng2022}. In finite-size samples, strain may give rise to valley polarized counterpropagating edge states, which could eventually lead to the detection of quantized valley Hall conductivities. Nevertheless, the lack of topological protection of these helical boundary modes makes their existence strongly dependent on the proper design of edge terminations and on the type of strain~\cite{Salerno2017}. Even more, short-range scattering can couple both valleys and lead to backscattering between the edge channels~\cite{Low2010}, making usual transport probes of the Fermi surface states very fragile to edge disorder and matching conditions. 

In this work, we demonstrate that a quantized valley Hall response can be directly measured by monitoring the variation of the particle density in the bulk, upon small variations of the applied strain; see Figure~\ref{Fig_1_story}. This method builds on the Widom-St\v{r}eda formula~\cite{Streda1982,Widom1982}, which relates (in its original form) the electrical Hall conductivity to a bulk density response: $\sigma_H=e \, \partial n_{\text{bulk}}/\partial B$, where $B$ denotes an applied magnetic field, $n_{\text{bulk}}$ is the particle density evaluated locally in the bulk and $e$ is the charge of the carriers. In the present framework of strained systems, we consider a \emph{pseudo-magnetic} field perturbation, as obtained by modifying the strength of the applied strain, so that the resulting density response directly reflects the underlying valley Hall conductivity. Importantly, this approach suggests that the quantized valley Hall response can be cleanly extracted from a bulk property, in sharp contrast with more standard non-local transport measurements.
This makes our proposal particularly appealing for synthetic lattice systems, as for engineered molecular lattices where the local density of states (LDoS) can be extracted via STM imaging~\cite{Gomes2012,Freeney2020, Swart2017,Drost2017, Khajetoorians2019,Polini2013},
or for the case of ultracold atoms in optical lattices, where the local particle density can be finely measured using the quantum gas microscope technique~\cite{Greiner2009, Kuhr2010, Greiner2015, Zwierlein2015, Gross2015, Kuhr2015,leonard2022realization}.

\section*{Results}

\begin{figure}[!b]
	\center
	\includegraphics[width=\columnwidth]{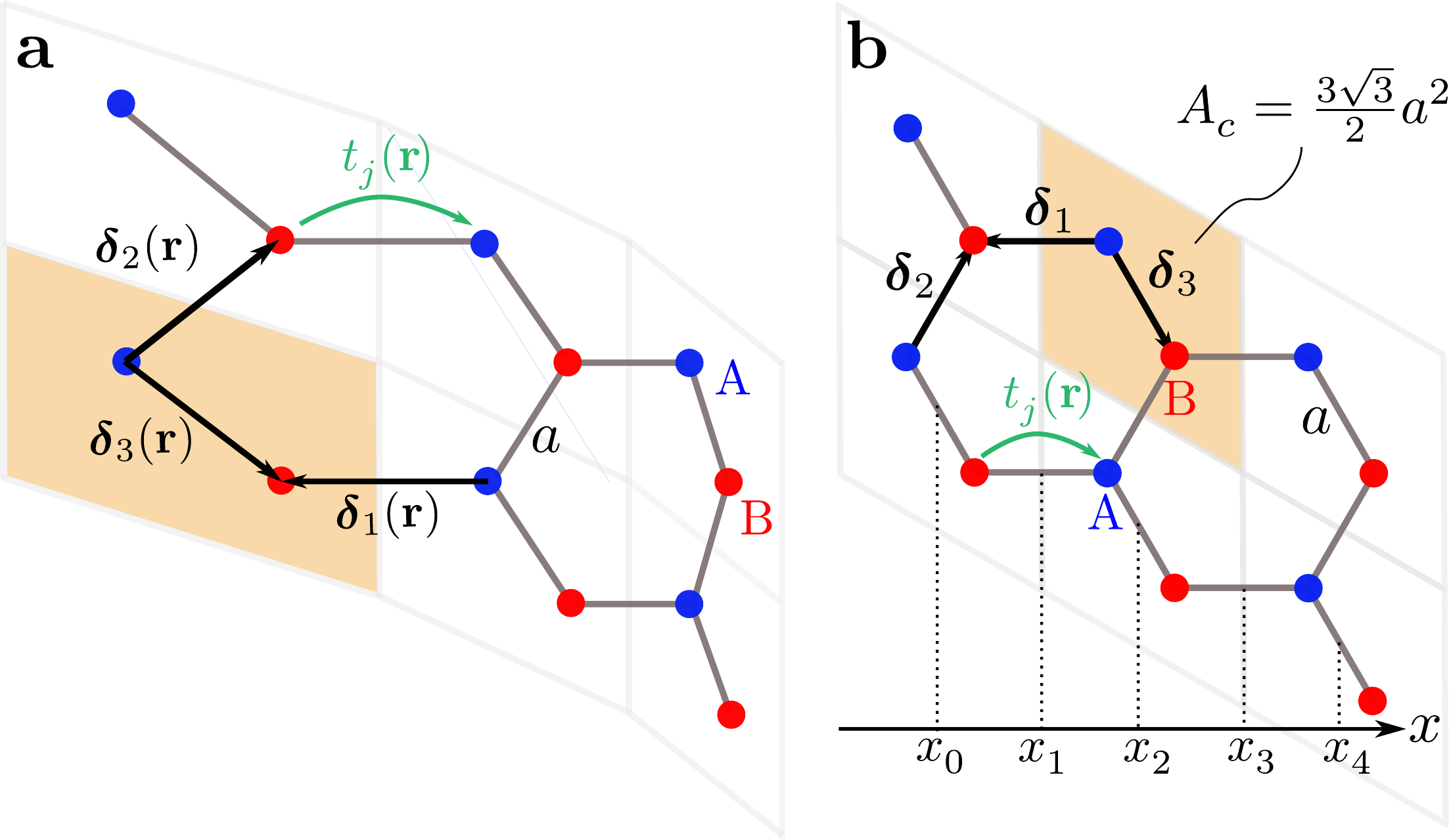}
	\caption{Honeycomb lattice with zig-zag terminations along the $x$-direction and $N_x=3$ primitive cells of area $A_c=3\sqrt{3}a^2/2$. The vectors connecting neighbouring sites are denoted as $\bm{\delta}_j$. (\textbf{a}) Strain is applied by  mechanically stretching the lattice. (\textbf{b}) Strain is directly imprinted on the tunneling amplitudes $t_j^{}(\bm r)$ without modifying the underlying crystalline structure of the lattice .}
	\label{fig:lattice}
\end{figure}
\begin{figure}[!t]
	\center
	\includegraphics[width=1.0\columnwidth]{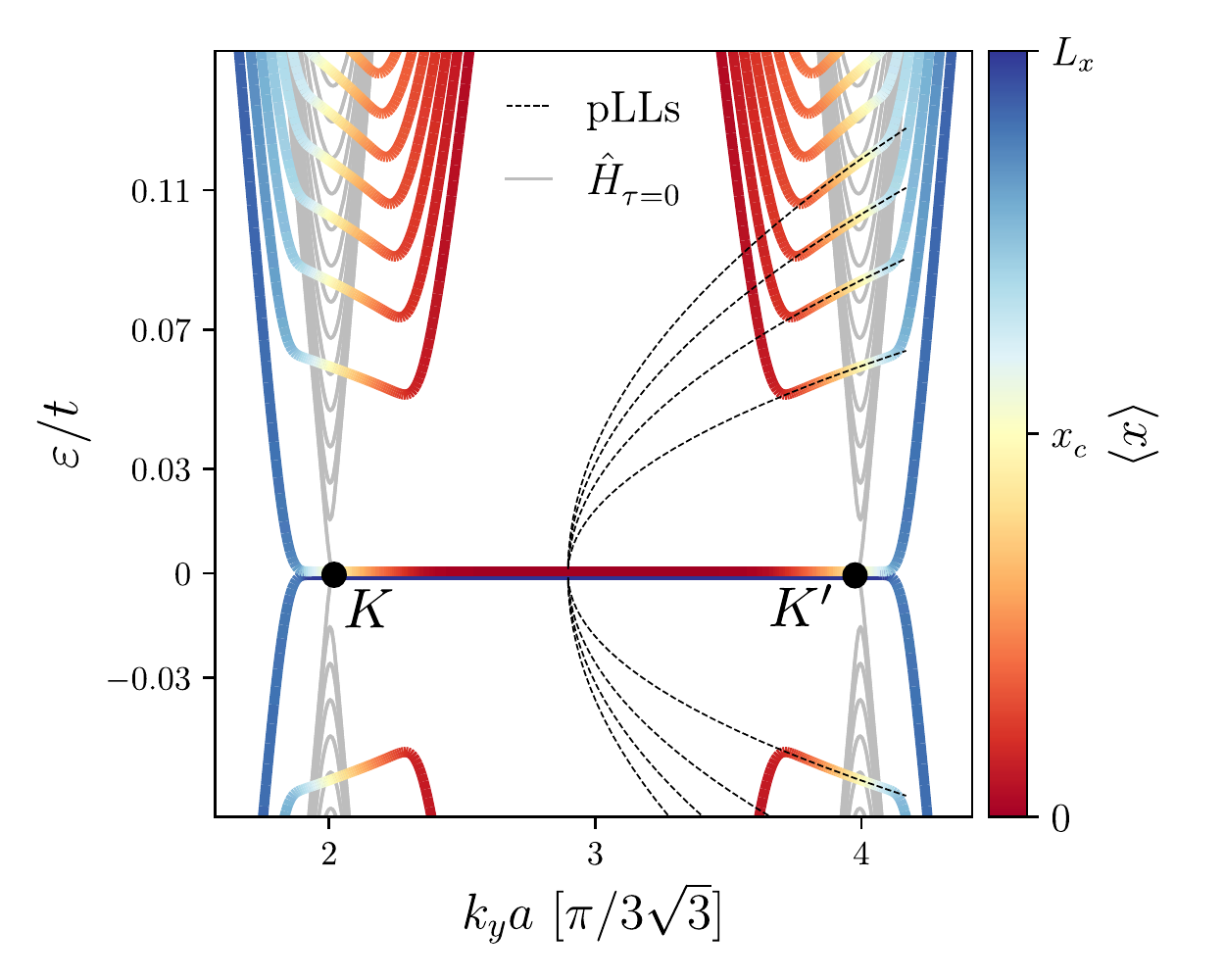}
	\caption{Energy spectrum of the Hamiltonian in Eq.~\eqref{H_strain} for $\tau=0$ (grey lines) and $\tau = 70.84\times 10^{-4}$ (colored lines). We have used $N_x=301$ cells along the $x$-direction and zigzag terminations. For finite strain, the color scale indicates the mean position $\langle x \rangle$ of each state. The black dashed lines represent the pseudo-Landau levels (pLLs) whose approximate dispersion is given by Eq.~\eqref{E_pLL}.}
	\label{fig:spectrum_strain}
\end{figure}

\textbf{Strained honeycomb lattices: the model.} The tight-binding Hamiltonian of a strained honeycomb lattice can be generically written as
\begin{equation}\label{H_strain}
\hat H^{} = - \sum_{\bm{r} \in \mathcal A,j} t_j(\bm{r})(\hat a_{\bm{r}}^\dagger \hat b^{}_{\bm{r}+\boldsymbol \delta_j(\bm{r})} + \text{h.c.}), \quad j \in \{1,2,3\},
\end{equation}
where the operators $\hat a^{}_{\bm{r}} (\hat a^\dagger_{\bm{r}})$ and $\hat b^{}_{\bm{r}+\bm{\delta}_j(\bm{r})} (\hat b^\dagger_{\bm{r}+\bm{\delta}_j(\bm{r})})$ are annihilation (creation) operators on A and B sublattices at position $\bm{r}$ and $\bm{r}+\bm{\delta}_j(\bm{r})$, respectively. In solid-state devices, such as mechanically stretched graphene sheets~\cite{Guinea2010a,Levy2010}, the hopping elements are essentially modified by distorting the lattice geometry, which is here encoded in the position of the atoms and in the set of space-dependent first nearest-neighbours vectors, which we defined as $\bm{\delta}_j(\bm{r})$ [see Fig.~\ref{fig:lattice}\textbf{a}]. This same strategy has been used in synthetically built systems, such as molecular graphene~\cite{Gomes2012,Freeney2020}, photonic~\cite{Rechtsman2013, Bellec2020, Jamadi2020} and acoustic~\cite{Abbaszadeh2017,Brendel2017,Wen2019} meta-materials. One of the main advantages of these setups is that stress configurations can be designed at will by simply engineering different lattice patterns. Other theoretical proposals have also analyzed the possibility of mimicking the physics of strained honeycomb lattices with ultracold atoms trapped in optical lattices~\cite{Alba2013,Tian2015,Jamotte2022,DiLiberto2022}. Interestingly, in some of those platforms, strain can be directly imprinted on the tunneling amplitudes without modifying the underlying crystalline structure of the lattice (see the Discussion Section). 
In that case, the vectors $\boldsymbol{\delta}_j(\bm{r})=\bm{\delta}_j$ are simply the pristine ones, namely $\boldsymbol{\delta}_1^{} = (-a,0)$, $\boldsymbol{\delta}_2^{} = (a/2,\sqrt 3a/2)$ and $\boldsymbol{\delta}_3^{} = (a/2,-\sqrt 3a/2)$ [see Fig.~\ref{fig:lattice}\textbf{b}], where $a$ is the lattice spacing. In the following, we present results for this simpler scenario, keeping in mind that the discussion can be easily generalized to the geometrically deformed lattice. We specifically work with \textit{uniaxial strain} along the $x$-direction, which is modeled with space-dependent tunneling amplitudes as
\begin{equation}\label{t_j}
\begin{split}
	t_j(x) = t\left(1+\tau \frac{x-x_c}{3a^2} |\hat{\mathbf x} \cdot \boldsymbol{\delta}_j |\right), \quad \hat{\mathbf x} = (1,0).
\end{split}
\end{equation}
Here, $t$ is a uniform tunneling amplitude, $\tau$ is the strain intensity and $x$ runs over discretized positions $x_l^{}$ located in the middle of a $\boldsymbol \delta_j^{}$-link between an A and a B site, as indicated in Fig.~\ref{fig:lattice}b. The position of the system's center is denoted by $x_\text{c}^{} = L_x^{}/2$, where $L_x^{}$ is the projected length of the ribbon along the $x$ direction. The parameters are kept in such a way that $\tau L_x/6a <1$, so that the hopping terms do not vanish across the entire sample. For this stress configuration, the Hamiltonian defined by Eq.~\eqref{H_strain} can be diagonalized for open boundary conditions along $x$ and periodic boundary conditions along $y$. The energy spectrum is shown in Fig.~\ref{fig:spectrum_strain}, where we present results for both the unstrained ($\tau=0$) and strained case ($\tau=70.84\times 10^{-4}$). We have used zig-zag terminations and $N_x=301$ cells along the $x$-direction, which correspond to a system size of $L_x=450.5\,a$. In the first case, the spectrum (grey lines) shows two well-defined Dirac valleys, with the gap closing points denoted as $\textbf{K}$ and $\textbf{K}'$. For $\tau>0$, (tilted) pseudo Landau levels are generated around each Dirac cone, and edge states arise, localized along the boundaries of the system. The lines of this spectrum have been colored based on the mean position $\left\langle x \right\rangle$ of each eigenstate. Since the spatial inhomogeneity in the hopping terms is not too strong ($\tau \ll 1$), the valley index is still preserved and it is possible to derive an effective Hamiltonian linearised around the Dirac points that incorporates the effect of strain via a minimal coupling term~\cite{Goerbig2011,Salerno2015a},
\begin{equation}\label{H_Dirac_A}
h^{\xi}_{\text A}(\bm{q}) = \hbar v_\text{F}^{} \left[(\xi q_y^{} - e A^{\xi}_y) \sigma_x^{} - (q_x^{}-eA^\xi_x) \sigma_y\right],
\end{equation}
where $v_F = 3ta/2\hbar$, $\mathbf A^{\xi}(x)=(0,A_y^\xi)=(\xi\hbar\tau/9ea^2)(x-x_c)\hat{\mathbf{y}}$ is the pseudo-vector potential, $\bm{q} = (q_x^{},q_y^{}) \equiv \bm{k} - \xi \mathbf K$ and $\xi$ indicates the valley, taking the value $ +1$ for $\mathbf K$ and $-1$ for $\mathbf K'$. The effective magnetic field at each valley is given by
\begin{equation}
\label{Buni}
\boldsymbol{B}_\tau^{\xi} = \boldsymbol{\nabla} \times \mathbf A^{\xi} =\xi \frac{\hbar\tau}{9e a^2} \hat{\mathbf{z}} = B_{\tau}^{\xi}\hat{\mathbf{z}}.
\end{equation} 
Note that it has opposite sign at each valley and is hence not a proper magnetic field but a pseudo-magnetic field, which does not break time-reversal symmetry. Consequently, the edge states in the sample are helical instead of chiral: counterpropagtaing modes emerge at each boundary, with the valley index determining the sign of their velocity, as clearly seen in the color code of Fig.~\ref{fig:spectrum_strain}.

The bulk energy spectrum consists of a set of discretized energy levels, which are not strictly flat, as Eq.~\eqref{H_Dirac_A} would predict. Indeed, there are higher-order terms, which have been neglected in the linear approximation that complicate the picture away from the Dirac points. The first correction to the constant pseudo-magnetic field model can be incorporated by considering that strain also induces a spatial dependence on the Dirac-Fermi velocity. Including this inhomogeneity, the pseudo-Landau levels acquire a dispersion in momentum space, which is approximately described by~\cite{Lantagne-Hurtubise2020}
\begin{equation}\label{E_pLL}
E_\nu(q_y) = \pm t \sqrt{\frac{\tau \nu}{2} \left(1-\xi \frac{3 q_y a}{2}\right)}, \quad \nu \in \mathbb N.
\end{equation} 
This analytical prediction is highlighted with black dashed lines in Fig.~\ref{fig:spectrum_strain}. We point out that this low-energy (effective) description is only valid close to the Dirac points $\mathbf K$ and $\mathbf K'$.

\textbf{The valley Hall response as a density response function.} The Widom-St\v{r}eda formula provides an insightful connection between the Hall conductivity of a two-dimensional gas $\sigma_H$ and the variation of its bulk particle density $n_{\textrm{bulk}}$ in response to the modification of an external magnetic field $B$. This relation states that, whenever the Fermi energy $\mu_F$ lies within a spectral gap, this transport coefficient can be obtained as
\begin{equation}\label{Streda}
\sigma_H^{} = e \frac{\partial n_{\textrm{bulk}}}{\partial B}\Bigg\rvert_{\mu_F^{}}.
\end{equation}
This formula, originally derived by St\v{r}eda within linear response theory~\cite{Streda1982}, was obtained independently by Widom using very general thermodynamic relations~\cite{Widom1982}. Interestingly, its validity holds for any insulating state of matter, including strongly-correlated ones~\cite{Repellin2020}. In the case of Chern insulators, such as quantum Hall states, Eq.~\eqref{Streda} can be used to predict the emergence of a quantized Hall response~\cite{Xiao2005,Umucalilar2008}.
Strained honeycomb lattices preserve time-reversal symmetry, so the Hall conductivity in these systems remains trivially equal to zero. Nevertheless, due to the explicit breaking of space-inversion symmetry, the quantum valley Hall effect can take place~\cite{Settnes2017}. When the Fermi energy is taken to be near the charge neutrality point, the valley Hall response $\sigma_\text{V}^{}$ can be defined as the difference of the contributions to the Hall conductivity at $\mathbf{K}$ and $\mathbf{K'}$, denoted $\sigma^K_\text{H}$ and $\sigma_\text{H}^{K'}$ respectively, i.e. 
\begin{equation}
\sigma_\text{V}^{} \equiv \sigma_\text{H}^{K}-\sigma_\text{H}^{K'}.    
\end{equation}
Our first goal is to adapt Eq.~\eqref{Streda} to the system under consideration, in order to probe the VHE via density response to strain variations. For $\tau\ll 1$, we can rely on the constant pseudo-magnetic field approximation discussed in the previous section to obtain the conductivity for each valley as 
\begin{equation}\label{Streda_strain_valley}
\sigma^{\xi}_\text{H} \simeq e\frac{\partial n^{\xi}_{}}{\partial B^{\xi}_\tau}\Bigg\rvert_{\mu_F} {\simeq}\,\,  \xi \left(\nu+\frac 12 \right)\sigma_0,
\end{equation}
where $n^{\xi}$ stands for the contribution of the $K$ ($\xi=+1$) or the $K'$ ($\xi = -1$) valley to the bulk particle density, $\nu$ is the index of the last occupied pLL and $\sigma_0 = e^2/h$ is the conductivity quantum. The last equality has been derived by performing an explicit calculation of $n^{\xi}$ with the analytical eigenstates obtained from Eq.~\eqref{H_Dirac_A} (see Appendix~\ref{AppendixA}). For uniaxial strain as the one of Eq.~\eqref{t_j}, these wavefunctions remain fairly close to the exact eigenstates near the center of the sample~\cite{Salerno2015a,Jamotte2022}, and can then be used to provide a good approximation to the particle density around the bulk of the system. Based on Eq.~\eqref{Streda_strain_valley}, the valley Hall response can be directly obtained in terms of the variation of the total bulk density $ n_\textrm{bulk}^{} = n^{K}_{} + n^{K'}_{}$, for varying strain intensity, as
\begin{eqnarray}
\notag
    \sigma_\text{V}^{}&\simeq&e \left(\frac{\partial n^K}{\partial  B^{K}_\tau} - \frac{\partial n^{K'}}{\partial B^{K'}_\tau}\right)\Bigg\rvert_{\mu_F}\\
\label{Streda_strain}
    &=& e \frac{\partial n_{\textrm{bulk}}}{\partial B_{\tau}^{K}}\Bigg\rvert_{\mu_F}\\
\notag
    &=& \frac{\partial \tilde{n}_{\textrm{bulk}}}{\partial \alpha_\tau^{}}\Bigg\rvert_{\mu_F} \sigma_0 \,\,{\simeq} \,\,(2\nu+1)\sigma_0.
\end{eqnarray}
Here, we defined 
\begin{equation}
\alpha_\tau^{} \equiv B_\tau^{K} A_c/\phi_0 ,   
\end{equation}
as the flux in a primitive cell of area $A_c = 3\sqrt 3a^2/2$ in units of the flux quantum $\phi_0 = h/e$ in the presence of a magnetic field $B_{\tau}^{K}$. In the last equality of Eq.~\eqref{Streda_strain}, $\tilde{n}_{\textrm{bulk}} = n_{\textrm{bulk}}A_c$ stands for the dimensionless particle density per cell in the bulk (i.e.~the number of particles within a unit cell illustrated in Fig.~\ref{fig:lattice}\textbf{b}).
Note that while usual experimental probes of the valley Hall conductivity rely on non-local (out-of-equilibrium) transport measurements, Eq.~\eqref{Streda_strain} provides an alternative approach, which only relies on locally testing the equilibrium particle density variations upon modifying the strength of strain. We stress that this approach relies on the low-energy Dirac model introduced in Eq.~\eqref{H_Dirac_A}, hence, it is valid in the regime $\tau \ll 1$.
\begin{figure}[!t]
	\center
	\includegraphics[width=1\columnwidth]{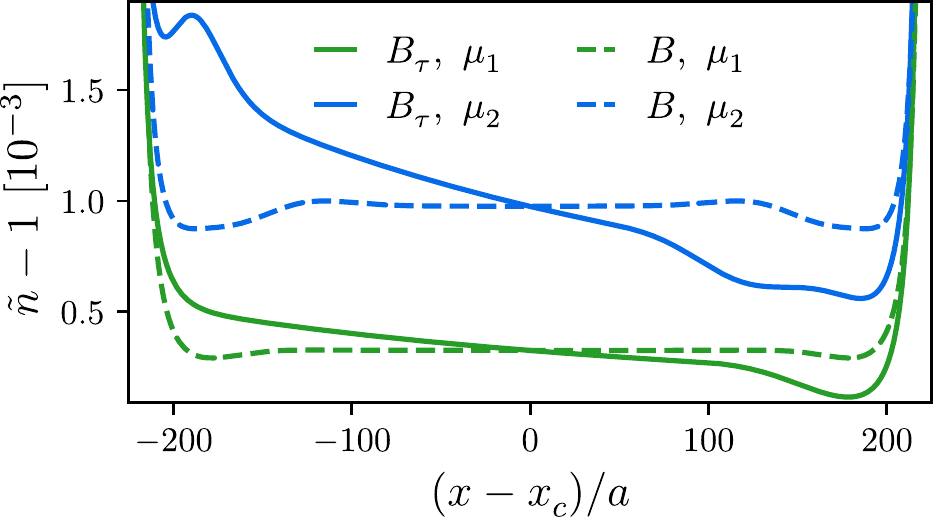}
	\caption{Particle density per cell for  the same parameters as in Fig.~\ref{fig:spectrum_strain} ($\alpha_\tau^{} = 3.255 \times 10^{-4}$) for a strained sample (solid line) and for a lattice with a homogeneous external magnetic field $B = B_\tau^{K}$ (dashed line). Green and blue colors correspond to chemical potentials $\mu_1^{} = 0.02\,t$ (first gap) and $\mu_2^{} = 0.068\,t$ (second gap), respectively. Note that we plot deviations of $\tilde{n}$ from unity, which are of the order of $10^{-3}$.}
	\label{fig:density_tot}
\end{figure}

In the case of uniaxial strain, the explicit dependence of the hopping amplitudes on the position brings a caveat to the problem: it makes the particle density to become position dependent, even deep into the bulk of the sample. This behaviour is explicitly shown in Fig.~\ref{fig:density_tot}, where we plot the dimensionless density of particles per cell 
\begin{equation}
\tilde{n}(x) = \sum_{\alpha=A,B} \tilde{n}(x_{\alpha}) = A_c\sum_{\alpha=A,B}  n(x_{\alpha}) ,
\label{ntildex}
\end{equation}
as a function of the cell position, which is here denoted as $x\equiv x_{2l}$ for $l=0,1,2\hdots$ (as defined in Fig.~\ref{fig:lattice}\textbf{b}). In Eq.~\eqref{ntildex}, the position of an $\alpha$-site within the unit cell at $x$ has been denoted as $x_{\alpha}$.
We show the behavior of this quantity for two different values of the chemical potential and the same parameters, as in Fig.~\ref{fig:spectrum_strain}. We also compare these densities with the ones of a honeycomb lattice in the presence of a homogeneous external magnetic field of strength $B=B_{\tau}^{K}$. Note that the difference between both models tends to zero at the center of the sample. Characteristic Friedel oscillations are clearly visible near the edges of the sample, until the density peaks due to the presence of the edge modes. As opposed to the constant magnetic field case, the strained lattice presents a clear asymmetry between the right and left boundaries. At the left of the sample, the hopping amplitude decreases with respect to its unstrained value [see Eq.~\eqref{t_j}], making the wavefunctions more localized and the density slightly higher than the one at $x=x_c$. The opposite behavior takes place at the right of the sample, where the density decreases with respect to the one at the center. Since the deviation is symmetrical with respect to this point, we can define a bulk particle density by averaging the densities $\tilde{n}(x)$ over a certain radius $r_{\textrm{bulk}}=L_{\textrm{bulk}}/2$ around $x=x_c$. In this way,
\begin{equation}
\tilde{n}_{\textrm{bulk}} = \frac{1}{N_{\textrm{bulk}}}\sum_{x\in\textrm{bulk}}\tilde{n}(x),
\label{n_tilde_bulk}
\end{equation}
where the bulk region corresponds to $x \in [x_c - r_{\textrm{bulk}}, x_c + r_{\textrm{bulk}}]$ and $N_{\textrm{bulk}}$ is the number of cells considered in the sum. If the bulk radius is small compared to the size of the system, the bulk density as defined above will remain quite close to the one that would be obtained from a constant magnetic field model, making the St\v{r}eda formulation of Eqs.~\eqref{Streda_strain_valley} and~\eqref{Streda_strain} still adequate. In particular, the valley Hall response may be obtained by averaging the density variations in the central bulk region as
\begin{equation}
\label{sigma_averaged}
\sigma_V = \frac{1}{N_{\textrm{bulk}}}\sum_{x\in \textrm{bulk}}\mathfrak{S}(x),
\end{equation}
where,
\begin{equation}
\label{local_marker}
\mathfrak{S}(x)=\sigma_0\frac{\partial \tilde{n}(x)}{\partial \alpha_{\tau}}\Bigg\rvert_{\mu_F}.    
\end{equation}
Note that $\mathfrak{S}(x)$ plays the role of a local marker in the problem: it provides a way to locally probe the valley Hall coefficient when properly averaged over $L_{\textrm{bulk}}$.

One must keep in mind that the valley Hall quantization is a property of an insulating bulk. In order to measure it, the Fermi energy has to lie in a region between two pLLs. A word of caution is in order here: since we are dealing with a finite-size sample in the presence of edge states, there are no true spectral gaps in the system. Nevertheless, the bulk particle density for sufficiently large systems ($L_{\textrm{bulk}} < L_x$) is expected to depend very weakly on the filling of the edge modes, and therefore the St\v{r}eda formulation should remain reasonably accurate, as discussed in the next section.

\begin{figure}[!t]
	\center
	\includegraphics[width=1.0\columnwidth]{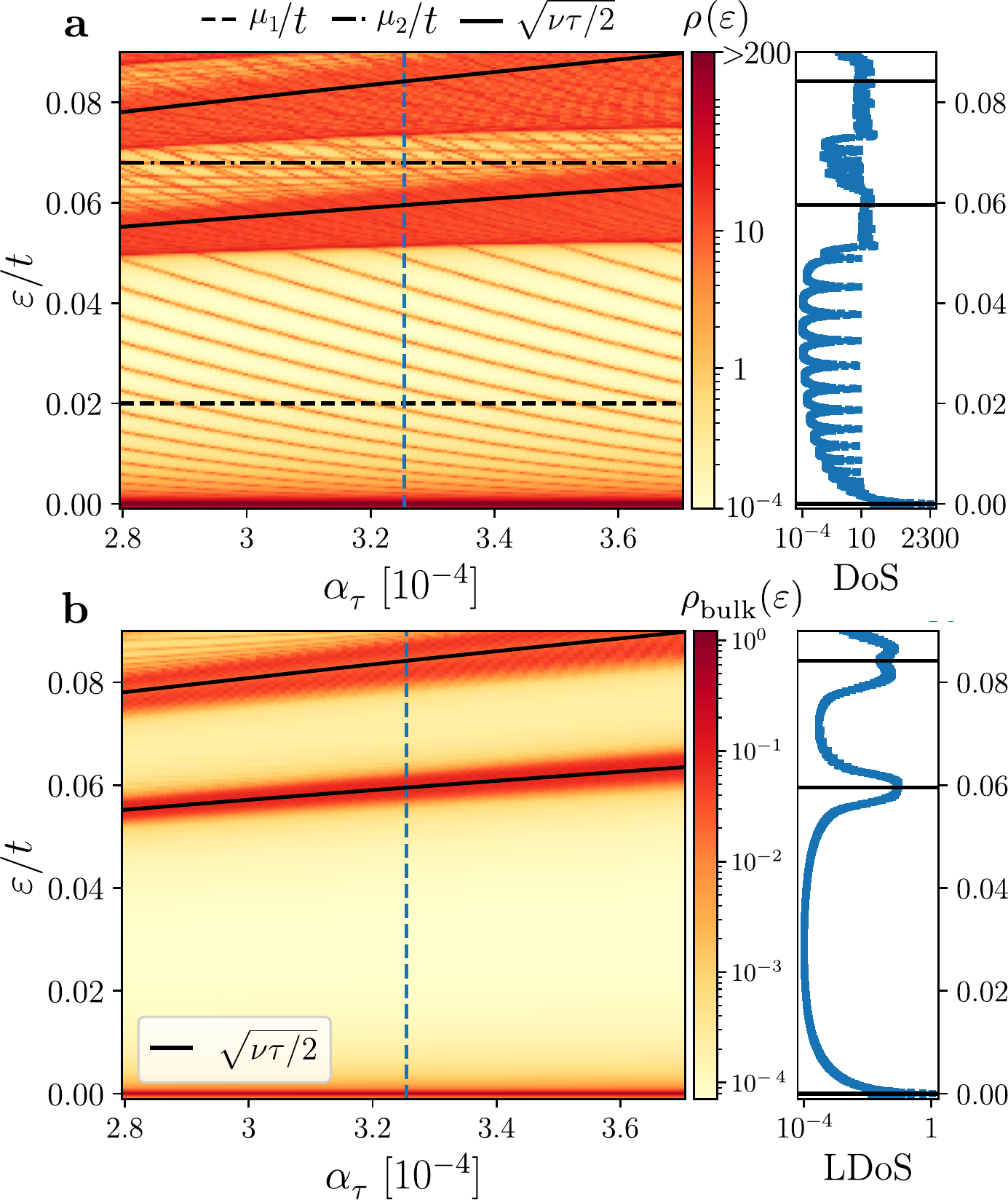}
	\caption{\textbf{a.} Total density of states $\rho(\varepsilon)$ in logarithmic scale as a function of energy $\varepsilon$ and pseudomagnetic flux $\alpha_\tau^{}$. The size of the sample is the same as in Fig.~\ref{fig:spectrum_strain}, namely $L_x = 450.5\,a$. The black dashed and dashed-dotted lines represent the chemical potentials $\mu_1^{} = 0.02t$ and $\mu_2^{} = 0.068 t$, respectively. \textbf{b.} Bulk density of states $\rho_{\textrm{bulk}}(\varepsilon)$ in logarithmic scale for $L_\textrm{bulk} = 36\,a$. In both panels, black solid lines identify the energy of the analytical pseudo-Landau levels at the Dirac points.  In the right panels, we show the corresponding DoS (or LDoS) for $\alpha^{}_\tau = 3.255 \times 10^{-4}$ indicated by the vertical dashed lines in the left panels.}
	\label{fig:Strain_Energy_alpha_sigma_spectra_alphas}
\end{figure}

\textbf{Spectral properties.} With the aim of determining the energy regions where the Widom-St\v{r}eda formula can be applied, it is instructive to study the spectral properties of the strained honeycomb lattice. We show in Fig.~\ref{fig:Strain_Energy_alpha_sigma_spectra_alphas}\textbf{a} the density of states (DoS) of the sample as a function of energy and $\alpha_{\tau}$, calculated as 
\begin{equation}
\rho(\varepsilon) = -\frac{1}{\pi}\textrm{Im}\textrm{Tr}[\hat{G}^r(\varepsilon)],
\end{equation}
with $\hat{G}^{r}(\varepsilon)=(\varepsilon + i\eta - \hat{H})^{-1}$ the retarded Green's function of the system. For reference purposes, the particular DoS for $\alpha_{\tau}^{}=3.255\,\times 10^{-4}$ is shown in the right panel of Fig.~\ref{fig:Strain_Energy_alpha_sigma_spectra_alphas}\textbf{a}, which corresponds to the value of strain ($\tau^{}=70.84\times 10^{-4}$) used to produce the spectrum of Fig.~\ref{fig:spectrum_strain}.
One clearly identifies a continuum of states representing the pseudo-Landau levels $\nu=0,1,2$, as well as a set of discrete modes, which stem from the edge states of the system. As opposed to the case of a non-strained honeycomb lattice in a real magnetic field, the pseudo-Landau levels for $|\nu|\geq 1$ have a certain width in energy due to their finite drift velocity. As a visual aid, we have included their analytical energy at the Dirac points in solid black lines -- see Eq.~\eqref{E_pLL} for $q_y^{} = 0$. The dependence of the DoS as a function of strain nicely reflects the spectral flow: when $\alpha_{\tau}$ increases, the discretized edge modes decrease in energy until they merge with the continuum of bulk states. This behavior is consistent with their wavefunction moving away from the hard-wall potential, while at the same time, becoming more localized in space (recall that the effective magnetic length $\ell_B^{} = \sqrt{\hbar/e B_{\tau}^{K}} = \sqrt{A_c^{}/2\pi  \alpha_\tau^{}}$).

The key quantity to evaluate the density variations through the sample  [Eqs.~\eqref{sigma_averaged} and~\eqref{local_marker}] is the local density of states per cell (LDoS), which may be obtained in terms of the retarded Green's function as
\begin{equation}
\rho(\varepsilon,x) = -\frac{1}{\pi}\sum_{\alpha=A,B}\textrm{Im}\langle x_\alpha | \hat{G}^{r}(\varepsilon)|x_\alpha\rangle,    
\end{equation}
where the sum runs over the two sublattice sites belonging to the cell at $x = x_{2l}^{}$.
Note that in order to obtain the local density of particles at each cell, this quantity must be integrated in energy up to the Fermi level,
\begin{equation}
\tilde{n}(x) = \int_{-\infty}^{\mu_F}\rho(\varepsilon,x)d\varepsilon.      
\end{equation}
The bulk particle density previously defined in Eq.~\eqref{n_tilde_bulk} can then be obtained as the average
\begin{equation}
\tilde{n}_{\textrm{bulk}} = \frac{1}{N_{\textrm{bulk}}}\sum_{x \in \textrm{bulk}}\int_{-\infty}^{\mu_F}\rho(\varepsilon,x)d\varepsilon = \int_{-\infty}^{\mu_F}\rho_{\textrm{bulk}}(\varepsilon)d\varepsilon.  
\end{equation}
Here, $\rho_{\textrm{bulk}}(\varepsilon) = \sum_{x\in \textrm{bulk}} \rho(\varepsilon,x)/N_\textrm{bulk}$ is nothing but the density of states projected onto this particular region. The valley Hall coefficient in Eq.~\eqref{Streda_strain} can be re-written in terms of this quantity as an integral over the Fermi sea
\begin{equation}
    \sigma_V = \sigma_0 \int_{-\infty}^{\mu_F}\frac{\partial\rho_{\textrm{bulk}}(\varepsilon)}{\partial \alpha_{\tau}}d\varepsilon.
\label{fixed_mu}
\end{equation}
In a finite-size sample, we thus expect to have a quantized result from Eq.~\eqref{fixed_mu} whenever $\rho_{\textrm{bulk}}(\mu_F)\simeq 0$.
We show in Fig.~\ref{fig:Strain_Energy_alpha_sigma_spectra_alphas}\textbf{b} the bulk density of states $\rho_{\textrm{bulk}}(\varepsilon)$ for a bulk region of width equal to $L_{\textrm{bulk}}=36\,a$ (centered around $x_c^{}$). This size of $L_{\textrm{bulk}}$ is of the order of the magnetic length for $\alpha_\tau^{} = 3.255 \times 10^{-4}$. We can clearly see that, for this particular bulk area, the contribution of the edge modes is negligible in the first two gaps between pLLs and slightly more relevant in the third gap (note the logarithmic color scale). This is in agreement with the edge states arising from the $\nu=2$ pLL being appreciably more delocalized than the ones originating from the $\nu=0$ and $\nu=1$ pLL. To avoid the presence of finite-size effects, we will then focus on the results for the first two gaps.

\begin{figure}[!t]
	\center
	\includegraphics[width=0.9\columnwidth]{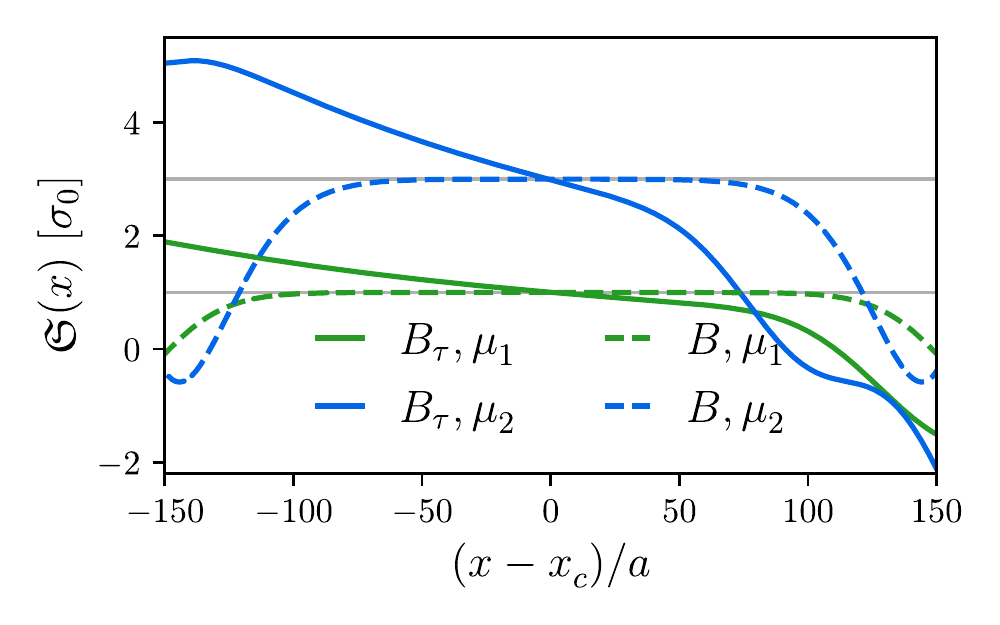}
	\caption{Spatial dependence of the kernel $\mathfrak{S}(x)$ for  the same parameters as in Fig.~\ref{fig:spectrum_strain} ($\alpha_\tau^{} = 3.255 \times 10^{-4}$) for a strained sample (solid lines) and for a lattice with a homogeneous external magnetic field $B = B_\tau^{K}$ (dashed lines). Green and blue colors correspond to chemical potentials $\mu_1^{} = 0.02\,t$ (first gap) and $\mu_2^{} = 0.068\,t$ (second gap), respectively.}
	\label{fig:cond_space}
\end{figure}

\textbf{A local probe for the valley Hall response.} Figure \ref{fig:cond_space} shows the local response $\mathfrak{S}(x)$ as a function of the position in the sample -- see Eq.~\eqref{local_marker} -- for two different chemical potentials: $\mu_1=0.02\,t$ in the first gap (green) and $\mu_2=0.068\,t$  in the second gap (blue). We compare the local response as obtained for a strained lattice (solid lines) and an unstrained lattice in a uniform magnetic field of magnitude $B=B_{\tau}^{K}$ (dashed lines). In this latter case, the local marker (also obtained with Eq.~\eqref{local_marker}) remains uniform in the bulk of the system and equal to the quantized integer value expected from the theory. This could be anticipated from Fig.~\ref{fig:density_tot}: a real magnetic field leads to plateaus in the bulk density profiles and, accordingly, in their variation with respect to the external flux. On the other hand, for the strained case, $\mathfrak{S}(x)$ presents a linear drift around the center of the system, naturally inherited from the density asymmetry resulting from the uniaxial strain we previously discussed. Deviations from this linearity arise as soon as edge effects become significant. These are more pronounced for the second than for the first gap, as the edge states from the former are more delocalized. Note that these local markers are equal at $x_\text{c}^{}$, confirming that the constant pseudo-magnetic field picture is accurate at the center of the sample. One also deduces from the linearity that discrepancies can be filtered away by simply averaging over an adequate radius $r_{\textrm{bulk}}=L_{\textrm{bulk}}/2$, as prescribed by Eq.~\eqref{sigma_averaged}. 
\begin{figure}[!t]
	\center
	\includegraphics[width=\columnwidth]{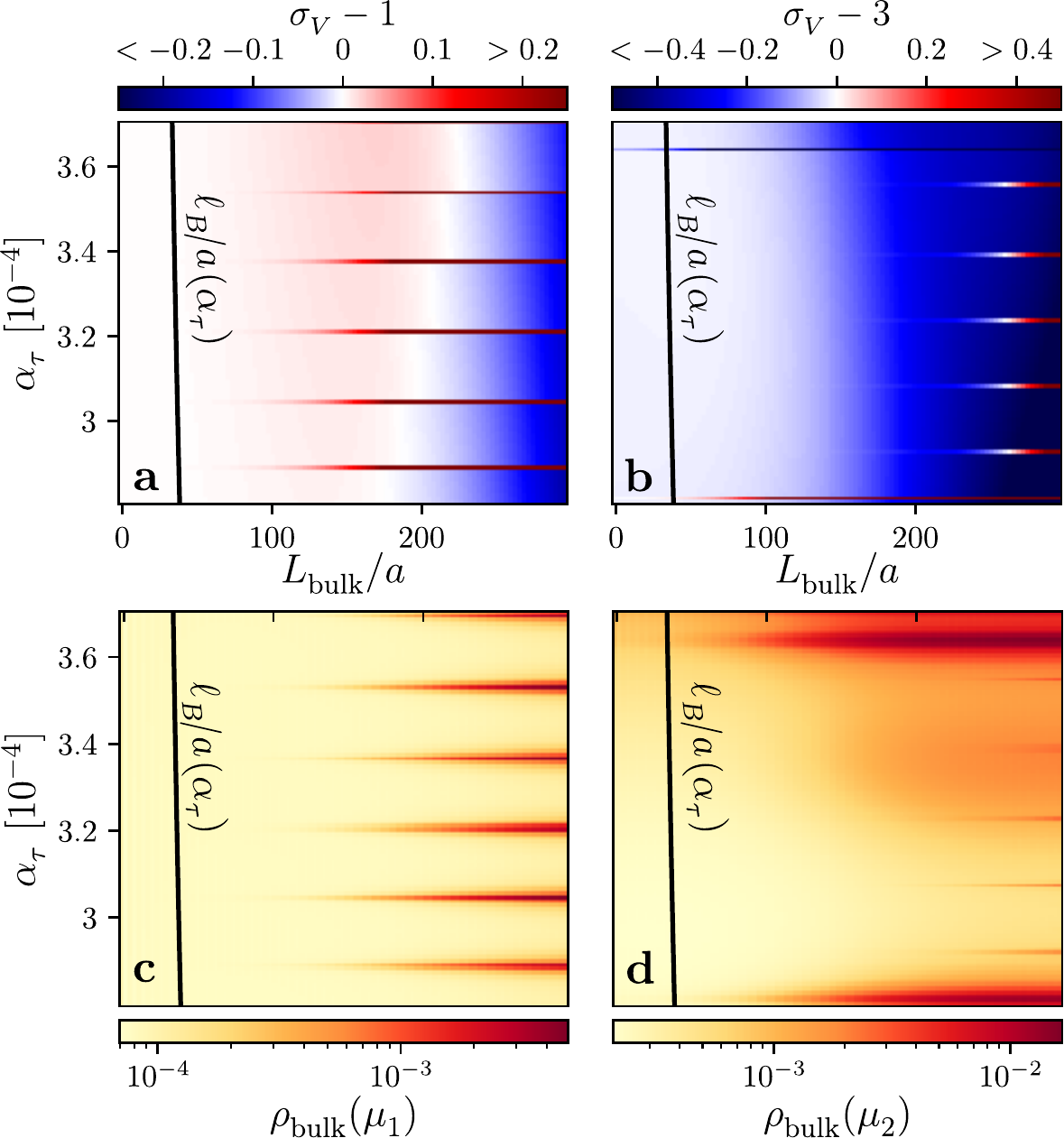}
	\caption{\textbf a-\textbf b. Valley Hall response obtained from Eq.~\eqref{sigma_averaged} as a function of the flux per plaquette $\alpha_{\tau}$ and the size of the bulk $L_{\textrm{bulk}}$ in a system of size $L_x^{} = 450.5 a$. The chemical potential has been fixed to $\mu_1=0.02\,t$ in the first and $\mu_2=0.068\,t$ in the second gap, respectively. The solid black lines represent the magnitude of the magnetic length $\ell_B^{}$ for each strain intensity. \textbf{c}-\textbf{d} Projected density of states $\rho_\textrm{bulk}(\mu_F)$ as a function of $\alpha_{\tau}$ and $L_{\textrm{bulk}}$ at the corresponding chemical potentials $\mu_\text{F}^{}=\mu_1$ and $\mu_\text{F}^{}=\mu_2$.}
	\label{fig:cond_Lbulk_tau_tot}
\end{figure}

\begin{figure}[!t]
	\center
	\includegraphics[width=1\columnwidth]{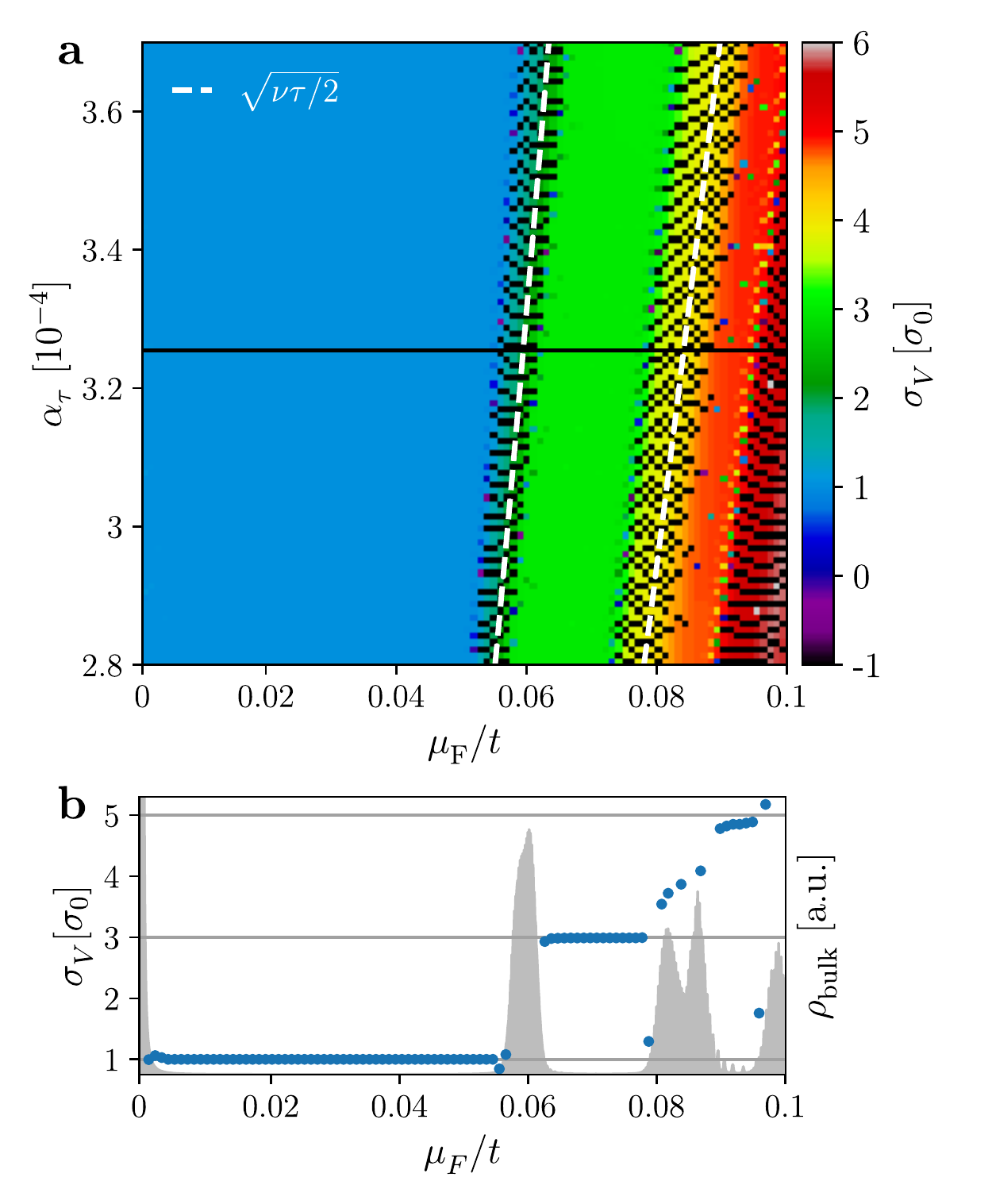}
	\caption{\textbf{a.} Valley Hall response obtained from Eq.~\eqref{sigma_averaged} as a function of $\alpha_{\tau}$ and $\mu_F$. Dashed white lines represent $\mu_F/t = \sqrt{\tau\nu/2}$ for $\nu=1$ and $\nu=2$. The size of the bulk region where we averaged the local marker is $L_{\textrm{bulk}}=36\, a$. \textbf{b.} (Blue dots) $\sigma_\text{V}^{}$ for $\alpha_{\tau}^{} =3.255 \times 10^{-4}$ (indicated by a solid black line in the upper panel). Results are restricted to the present window for clarity. (Grey area) Local density of states $\rho_\text{bulk}^{}$ from Fig.~\ref{fig:Strain_Energy_alpha_sigma_spectra_alphas}\textbf{b}.}
	\label{fig:cond_mu}
\end{figure}
In order to properly determine a reasonable bulk size to perform the average, we show in Fig.~\ref{fig:cond_Lbulk_tau_tot} the valley Hall response as obtained from Eq.~\eqref{sigma_averaged} as a function of $\alpha_{\tau}$ and $L_{\textrm{bulk}}^{}$ for $\mu_\textrm{F}^{} = \mu_1$ (panel $\textbf{a}$) and $\mu_\textrm{F}=\mu_2^{}$ (panel $\textbf{b}$). In both cases, the bulk density response to strain variations shows a remarkably quantized value ($\sigma_V^{} \simeq 1$ and $\sigma_V^{} \simeq 3$, respectively) when averaging over $L_{\textrm{bulk}}\lesssim 100 a$. Noticeable deviations from these integer values occur at very specific values of $\alpha_{\tau}$, appearing as horizontal stripes. For these values of strain, one of the discretized modes that stem from the edge states crosses the Fermi level, as can be seen in Fig.~\ref{fig:Strain_Energy_alpha_sigma_spectra_alphas}\textbf{a}. This is evidenced in Fig.~\ref{fig:cond_Lbulk_tau_tot}$\,$\textbf{c},\textbf{d}, where we show the bulk density of states at the Fermi energy $\rho_\text{bulk}^{}(\mu_\text{F}^{})$ as a function of $\alpha_\tau^{}$ and $L_\text{bulk}^{}$ for the same chemical potentials as the ones used for the upper panels.
If the size of $L_{\textrm{bulk}}$ is sufficiently large, the density of states projected onto the selected region becomes finite, breaking down the insulating character of the portion of the system that is being probed. Indeed, $\rho_\text{bulk}^{}(\mu_\text{F}^{})$ also presents horizontal stripes at the same values of $\alpha_\tau^{}$ as in Fig.~\ref{fig:cond_Lbulk_tau_tot}\textbf{a} and Fig.~\ref{fig:cond_Lbulk_tau_tot}\textbf{b}, reflecting an increase of the contribution from the edge states. More generally, a quantized valley Hall coefficient can be measured via bulk density response to strain variations as long as $\rho_\text{bulk}^{}(\mu_\text{F}^{}) \simeq 0$. When finite-size effects become appreciable, $\rho_\text{bulk}^{}(\mu_\text{F}^{})$ increases, leading to deviations of the response from the quantized integer values predicted by the theory. We thus conclude from Fig.~\ref{fig:cond_Lbulk_tau_tot} that, within our range of parameters, any  $L_\text{bulk}^{} \lesssim 100a$ is a suitable bulk size for probing the quantum valley Hall effect in the first couple of gaps for a sample of size $L_x^{} = 450.5 \, a$. 

In  Fig.~\ref{fig:cond_mu}\textbf{a}, we plot the valley Hall response as obtained from Eq.~\eqref{sigma_averaged} as a function of both the chemical potential $\mu_F$ and $\alpha_{\tau}$, so as to display a full scan of the valley Hall fan diagram. Here, we have chosen an average region of size $L_{\textrm{bulk}}=36\,a$. As a guide to the eye, we include dashed white lines to identify whenever the Fermi energy is equal to the analytical $\nu$-th pLL energy, i.e. $\mu_F/t = \sqrt{\tau\nu/2}$. We can clearly see the formation of plateaus in all the regions where the filling fraction of the pseudo-Landau levels remains constant. Quantization breaks down as soon as the bulk becomes metallic. This is best illustrated in Fig.~\ref{fig:cond_mu}\textbf{b}, where we show a specific cut of the upper panel for $\alpha_{\tau}= 3.255 \times 10^{-4}$. The shaded grey area represents the bulk density of states $\rho_{\textrm{bulk}}(\mu_F)$. Whenever the region being probed becomes incompressible ($\partial\rho_{\textrm{bulk}}/\partial \mu_F^{}\simeq 0$), a robust valley Hall plateau occurs. On the other hand, as soon as $\rho_{\textrm{bulk}}(\mu_F)$ becomes finite, the density response to strain variations becomes erratic. 

\textbf{Trigonal strain in a finite hexagonal flake.} Up to this point, we particularized our analysis to large system sizes, as made possible by using uniaxial strain along one direction and periodic boundary conditions along the other. It is worth asking whether these results would hold for smaller samples with different strain configurations or edge terminations. In this section, we explore a more realistic geometry where open boundary conditions are imposed along the entire system. This is well suited for the implementation of trigonal strain, which we model with a space-dependent tunneling amplitude of the form
\begin{equation}
\label{hopping_triaxial}
    t_j(\bm{r}) = t\left(1 + \tau \frac{(\bm{r}-\bm{r}_c)\cdot\bm{\delta}_j}{3 a^2}\right),
\end{equation}
where $\bm{r}_c$ denotes the position of the system's center. This particular strain leads to an effective gauge potential of the form $e\bm{A}^{\xi}(\bm{r}) = \xi \hbar \tau/3 a^2\left[(y-y_c)\hat{\mathbf{x}} - (x - x_c)\hat{\mathbf{y}}\right]$ \cite{Salerno2017}. The corresponding pseudo-magnetic field at each valley is consequently given by
\begin{equation}
\bm{B}_{\tau}^{\xi}= \bm{\nabla}\times \bm{A}^{\xi} = -\xi\frac{2 \hbar \tau}{3 e a^2}\hat{\mathbf{z}}.
\end{equation}
Note that with this convention, aside from changing its magnitude, the sign of the pseudo-magnetic field is opposite at each of the valleys with respect to the previously analyzed case [see Eq.~\eqref{Buni}]. In this regard, we already anticipate a bulk density response to strain variations, which should be negative (positive) for $\tau>0$ ($\tau<0$), when taking a chemical potential $\mu_F >0$ within a gap between pLLs.
The triaxial deformation is known to produce pLLs that are non-dispersive, as opposed to the tilted ones obtained with the previously uniaxially-stretched lattices. We then expect to observe particle densities with a plateau-like behavior near the bulk of the system, which could lead to an improvement of the spatial homogeneity of the local valley marker $\mathfrak{S}(\bm r)$. Furthermore, these highly degenerate pLLs lead to larger bulk spectral gaps that could help detecting a more precise quantized response for higher filling fractions.
\begin{figure}[!t]
	\center
	\includegraphics[width=\columnwidth]{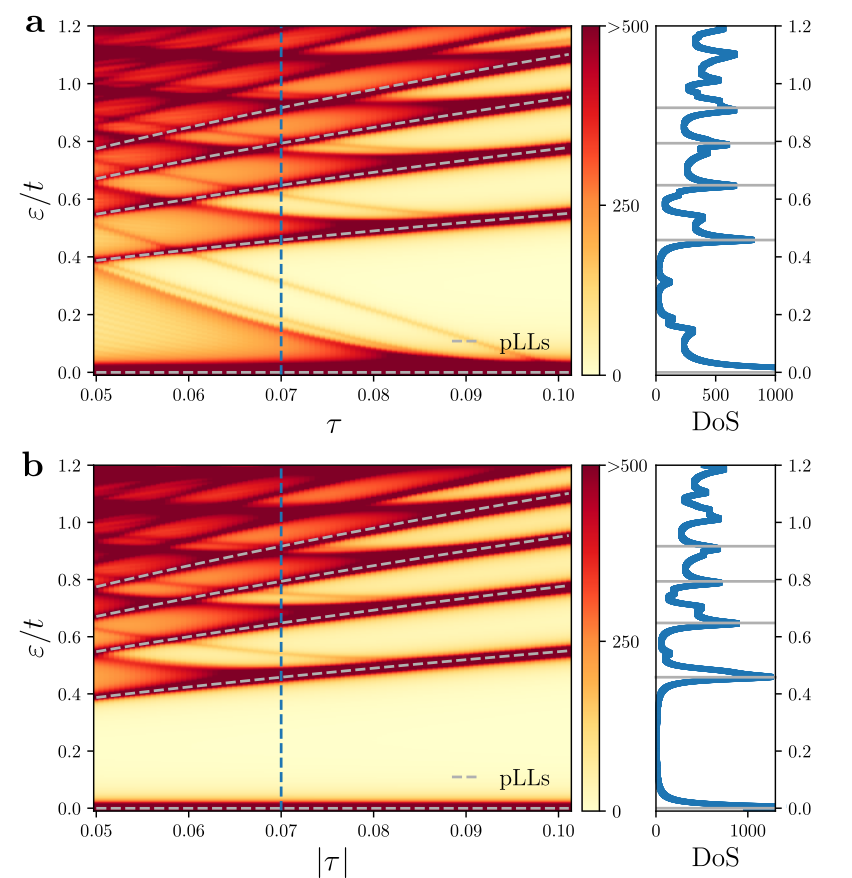}
	\caption{Total density of states $\rho(\varepsilon)$ in the hexagonal flake for (\textbf{a}) $\tau >0$  and (\textbf{b}) $\tau<0$. Dashed grey lines indicate the energy of the analytical pseudo-Landau levels $E_{\nu} = \pm v_F \sqrt{2\hbar e|\bm{B}_{\tau}^{\xi}|\nu}=\pm t \sqrt{3|\tau|\nu}$ for $\nu=0,1,2,3,4$. In the right panels, we show a specific cut of the left panels (marked with a vertical dashed blue line) for $|\tau| = 0.07$. Solid grey lines denote the analytical pseudo-Landau levels.}
	\label{dosflake}
\end{figure}
\begin{figure*}[!t]
	\center
	\includegraphics[width=\textwidth]{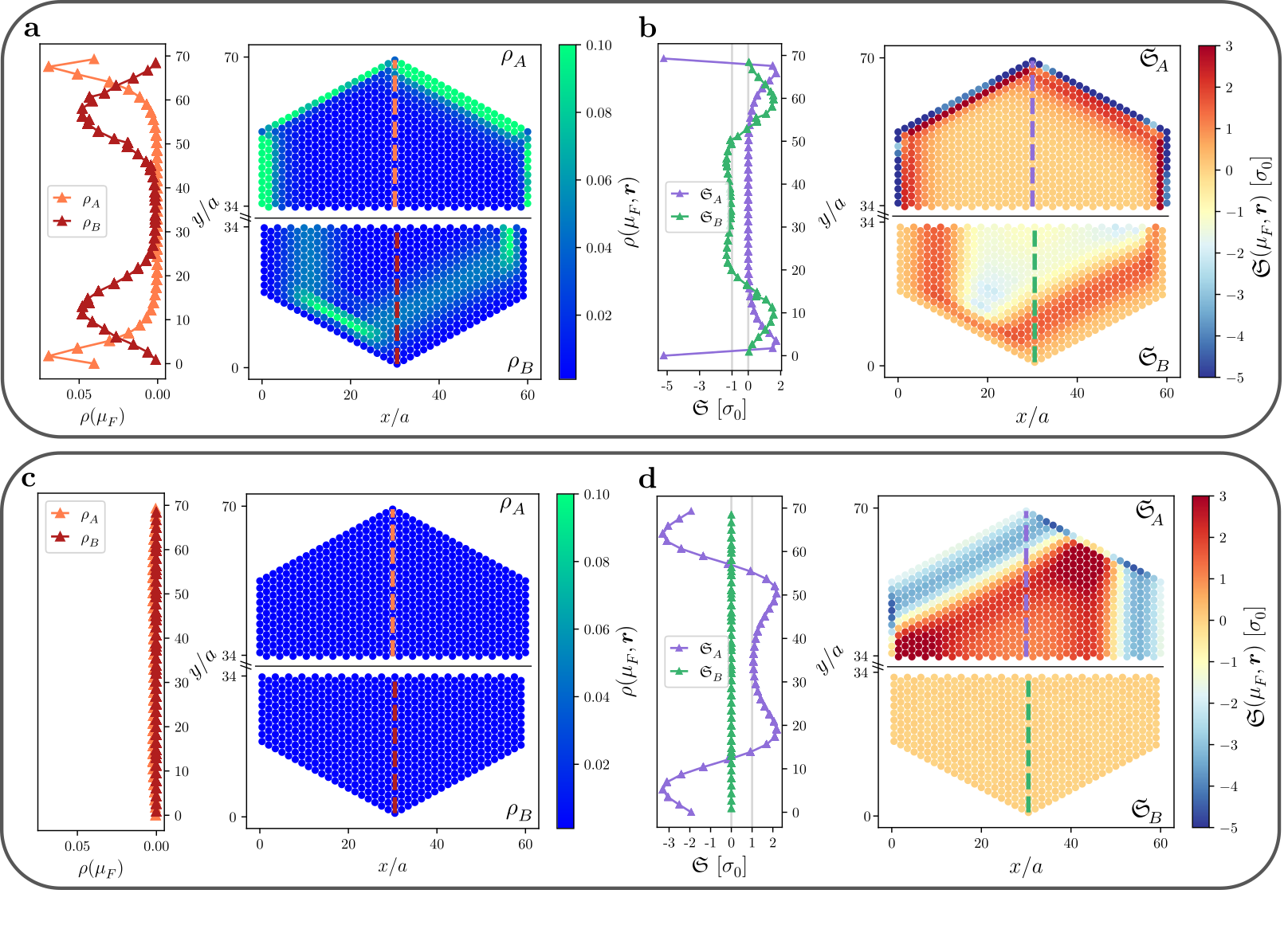}
	\caption{Panels \textbf{a} and \textbf{c} show the local density of states discriminated by sublattice $\rho_{\alpha}(\varepsilon,\bm{r})$ [see Eq.~\eqref{rho_sublattice}] for $\tau>0$ and $\tau<0$, respectively. The energy has been taken to be $\varepsilon=\mu_F=0.1\,t$ and $|\tau|=0.07$. The terminations have been chosen so as to alternate between bearded and zigzag-like edges. We exploit the mirror symmetry of the flake with respect to the $y=y_c$ axis to plot separately half of the A-sublattice (the upper part) and half of the B-sublattice (the lower part). A corresponding cut of the panels along $\bm{r}=x_c\hat{\mathbf{x}}+y\hat{\mathbf{y}}$ is shown to the left, where we have restored the entire spatial distribution. Panels \textbf{b} and \textbf{d} show the corresponding local marker $\mathfrak{S}_{\alpha}(\bm{r})$ differentiated by sublattice [see Eq.~\eqref{salpha}].} 
	\label{hexagonal_flakes}
\end{figure*}
We numerically study a triaxially-stretched flake of a honeycomb lattice, which is shaped in the form of a hexagon, hence preserving the trigonal symmetry. The terminations have been chosen so as to alternate between bearded and zigzag-like edges \cite{Salerno2017}. In this way, the perimeter of the hexagon can be built with sites belonging to a single sublattice (without loss of generality, we choose them to be A sites). Interestingly, depending on the sign of the hopping variations (namely $\tau$ being positive or negative) these types of flakes can support, or not support, helical edge states in the gap between the 0-th and 1-st pLL. As thoroughly described in Ref.~\cite{Salerno2017}, this essentially depends on the wavefunction of the $\nu=0$ pLL being localized on the B-sublattice ($\tau>0$ within our convention) or on the A-sublattice ($\tau<0$). In the former case, the zero energy pLL can be mixed with the non-propagating edge modes that live on the A-sublattice, generating dispersive helical edge states, while in the latter case the chiral symmetry forbids this to happen. We will present results for both cases and show that the density response $\mathfrak{S} (\bm r)$ that we propose as a valley Hall marker is quantized regardless of the edge states being present or not in the first gap.

We show in Fig.~\ref{dosflake} the flake's total density of states $\rho(\varepsilon)$ as a function of energy and the strain strength for the case $\tau>0$ (upper panel) and $\tau<0$ (lower panel). A pseudo-Landau level structure is revealed for both cases. The bulk spectrum is reasonably well described by the low-energy approximate model, namely $E_{\nu} = \pm v_F \sqrt{2\hbar e|\bm{B}_{\tau}^{\xi}|\nu}=\pm t \sqrt{3|\tau|\nu}$, which is displayed by grey dashed lines. As already anticipated, for positive $\tau$, edge states emerge in the first gap, while for negative $\tau$ the density of states remains exactly zero within that range of energies, indicating the absence of propagating helical modes. We also note that boundary modes are present in all gaps between higher pseudo-Landau levels ($\nu>0$), in accordance with the criterion developed in Ref.~\cite{Salerno2017}. This behavior can be better captured by analyzing the local density of states at a given energy, which we here define for each sublattice site as
\begin{equation}
\rho_{\alpha}(\varepsilon,\bm{r}) = -\frac{1}{\pi}\textrm{Im}\langle \bm{r}_{\alpha} | \hat{G}^{r}(\varepsilon)|\bm{r}_{\alpha}\rangle,
\label{rho_sublattice}
\end{equation}
where $\alpha=A,B$. In Fig.~\ref{hexagonal_flakes}~\textbf{a} and \textbf{c}, this quantity is plotted for both positive and negative strain variations, respectively. We have chosen an energy within the first gap $\varepsilon=\mu_F^{} = 0.1\,t$ and $|\tau|=0.07$. For this large pseudo-magnetic field, the magnetic length is equal to $\ell_B\simeq 4.63\,a$. We exploit the mirror symmetry of the flake with respect to the $y=y_c$ axis to plot separately half of the A-sublattice (the upper part) and half of the B-sublattice (the lower part). The missing pieces can be simply obtained by reflecting the corresponding portions with respect to the horizontal line that divides the sample into two halves.
In Fig.~\ref{hexagonal_flakes}\textbf{a}, delocalized edge modes surrounding the entire perimeter of the flake are clearly visible. A corresponding cut along $\bm{r}=x_c\hat{\mathbf{x}}+y\hat{\mathbf{y}}$ is shown to the left of the figure, where we have restored the entire spatial distribution. 
In Fig.~\ref{hexagonal_flakes}\textbf{c} the LDoS remains strictly zero along the entire flake, in consistency with the absence of edge states for this sign of $\tau$.

In the right-hand side of Fig.~\ref{hexagonal_flakes}, we present the local valley Hall marker for $\tau>0$ (\textbf{b}) and $\tau<0$ (\textbf{d}). The chemical potential has been chosen to be $\mu_F = 0.1\,t$. The density response to strain variations is discriminated by sublattice,  that is to say, we plot individually
\begin{equation}
\mathfrak{S}_{\alpha}(\bm{r}) = \sigma_0 \frac{\partial \tilde{n}_{\alpha}(\bm{r})}{\partial \alpha_{\tau}}\Bigg\rvert_{\mu_F}=\sigma_0 \int_{-\infty}^{\mu_F}\frac{\partial \rho_{\alpha}(\varepsilon,\bm{r})}{\partial \alpha_{\tau}}d\varepsilon,
\label{salpha}
\end{equation}
where $\tilde{n}_{\alpha}(\bm{r})$ is the dimensionless particle density at position $\bm{r}_{\alpha}$ and $\alpha=A,B$. We can clearly see that the response near the bulk of the flake is quantized in both scenarios: when the system supports or lacks edge states. Even more, the response is quantized separately on each sublattice (see Appendix~\ref{AppendixA}). Due to the particle-hole symmetry of the problem, the contribution to the total particle density of the half-filled system must remain inert to strain variations. In this sense, the bulk valley marker for chemical potentials within the first gap can be analyzed by only considering the modifications of the particle density of the 0-th pLL, which is localized either on the B-sublattice for $\tau>0$ or on the A-sublattice for $\tau<0$. The sublattice polarization of the bulk pLL leads to a sublattice polarization of the marker and explains why $\mathfrak{S}_{A}(\bm{r})\simeq 0$ and $\mathfrak{S}_{B}(\bm{r})\simeq -1$ for $\bm{r}$ around $\bm{r}_c$ in Fig.~\ref{hexagonal_flakes}\textbf{b}, while in Fig.~\ref{hexagonal_flakes}\textbf{d} the opposite behavior takes place, namely $\mathfrak{S}_{A}(\bm{r})\simeq 1$ and $\mathfrak{S}_{B}(\bm{r})=0$. Note that, in this last case, the response of the B sites is strictly zero in the entire sample: due to the absence of edge modes and the localization of the 0-th pLL on the A sites, the B sublattice remains completely half-filled for a chemical potential within the first gap meaning that $\tilde{n}_B(\bm{r})=1/2$ for every $\bm{r}$ and hence $\partial\tilde{n}_B(\bm{r})/\partial \alpha_{\tau}\rvert_{\mu_F}=0$. We stress that the plateau-like behavior of the local marker in Fig.~\ref{hexagonal_flakes} is here inherited from the fact that the particle densities of triaxally strained lattices are more uniform around the bulk than the tilted ones previously obtained for the uniaxial configuration. This can be seen in the left panel of Fig.~\ref{Fig_1_story}, where we plotted the adimensional particle density per cell $\tilde n(\bm r)$ for the same parameters as in Fig.~\ref{hexagonal_flakes}$\mathbf{d}$. 

\begin{figure}[!t]
	\center
	\includegraphics[width=\columnwidth]{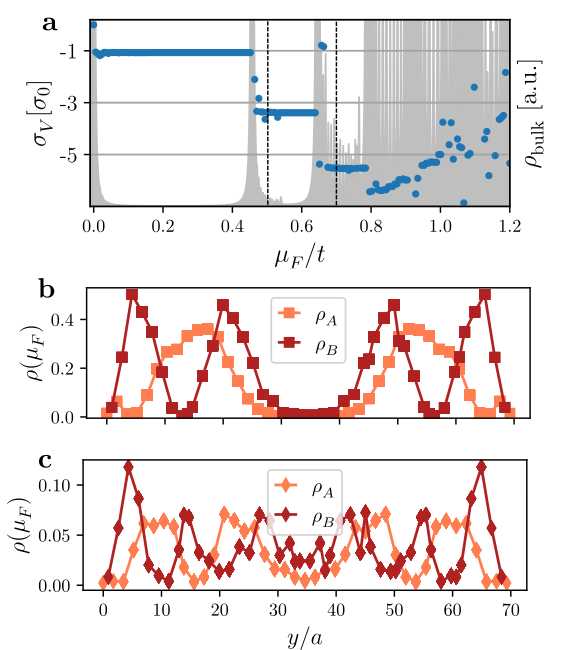}
	\caption{\textbf{a.} Valley Hall response $\sigma_V$ (blue points) obtained from averaging the local marker per cell $\mathfrak{S}(\bm{r}) = \mathfrak{S}_{A}(\bm{r})+\mathfrak{S}_{B}(\bm{r})$ in a bulk region of size $L_{\textrm{bulk}}=5\,a$ as a function of $\mu_F$. The grey shaded area shows the density of states projected onto the bulk region at the Fermi level $\rho_\text{bulk}^{}(\mu_F)$. Panels \textbf{b} and \textbf{c} show the local density of states $\rho(\mu_F,\bm{r})$ along $\bm{r} = x_c\hat{\mathbf{x}} + y\hat{\mathbf{y}}$ for chemical potentials $\mu_F=0.52\,t$ and $\mu_F = 0.7\,t$, respectively, indicated by black dashed lines in \textbf{a}.}
	\label{cond_mu_rhobulk}
\end{figure}
The valley Hall coefficient $\sigma_V^{}$ may be obtained by following the same procedure as before, specifically, averaging the local marker per cell $\mathfrak{S}(\bm{r})=\mathfrak{S}_{A}^{}(\bm{r})+\mathfrak{S}_B^{}(\bm{r})$ over a reasonable bulk radius $r_{\textrm{bulk}}=L_{\textrm{bulk}}/2$ [see Eq.~\eqref{sigma_averaged}]. Since the bulk response in Fig.~\ref{hexagonal_flakes} remains fairly homogeneous and quantized over a magnetic length around the center of the system, we take $L_{\textrm{bulk}}=5\,a$, which is of the order of the magnetic length $\ell_B^{}$ for $\tau = 0.07$. The averaged response as a function of chemical potential is shown in Fig.~\ref{cond_mu_rhobulk}$\mathbf{a}$ for the case of $\tau = 0.07 > 0$. We also include with a shaded grey area the density of states projected onto the bulk region being probed. The first couple of Lorentzian-shaped peaks represent the bulk pLL states that, as opposed to the uniaxially strained case, are here well defined in energy. The valley Hall coefficient remains quantized to a good degree for chemical potentials within the first gap. Nevertheless, finite-size effects already become appreciable for gaps between higher pLLs. In Fig.~\ref{cond_mu_rhobulk}~$\mathbf{b}$ and $\mathbf{c}$ we show the behavior of the local density of states at chemical potentials within the second and third gaps ($\mu_F = 0.52\,t$ and $\mu_F^{} = 0.7\,t$), which are indicated by black dashed lines in  Fig.~\ref{cond_mu_rhobulk}\textbf{a}. Note that the number of nodes of the LDoS differentiated by sublattice nicely reflects the polynomial hierarchy of the Dirac eigenspinors subject to a pseudo-magnetic field. In contrast to the case of $\mu_F = 0.1\,t$ (Fig.~\ref{hexagonal_flakes}$\mathbf{a}$), the edge modes for these energies are much more delocalized and hence lead to a breakdown of the insulating character of the bulk portion of the system that is being probed. As already discussed in the previous section, the equivalence between the valley Hall coefficient and the density variations strongly relies on probing an incompressible region. In this sense, the deviations from the expected quantized result taking place for chemical potentials within the second and third gaps can be partly attributed to a finite $\rho_{\textrm{bulk}}$ at the Fermi level. On the other hand, for these large values of strain, the low-energy model fails to describe correctly the higher order bulk pLLs, which leads to another natural source of discrepancy with the analytical prediction. 

\begin{figure*}[!t]
	\center
	\includegraphics[width=1\textwidth]{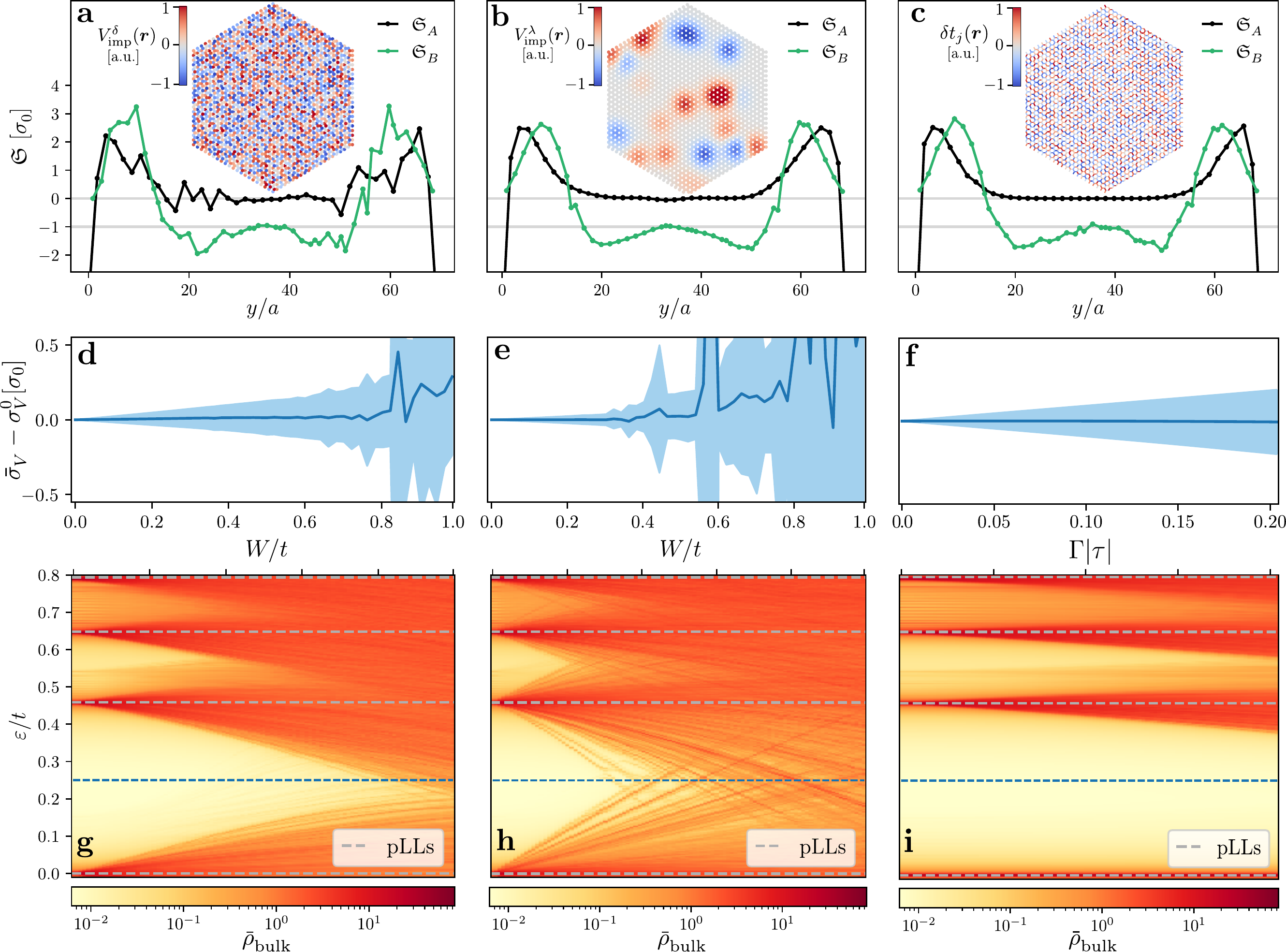}
 \caption{Effect of lattice imperfections on the valley Hall coefficient of the triaxially strained flake with $\tau = 0.07>0$ ($\ell_B^{} = 4.6a$) and $\mu_F =0.25\,t$. \textbf{First column}: Short-range on-site potential $V_{\textrm{imp}}^{\delta}(\bm{r})$ with $N_{\textrm{imp}}=N_{\textrm{tot}}$. \textbf{Second column}: Gaussian impurities described by $V_{\textrm{imp}}^{\lambda}(\bm{r})$ with an impurity concentration of $N_{\textrm{imp}}/N_{\textrm{tot}}=0.01$ and $\lambda = 3\,a$. \textbf{Third column}: Off-diagonal bond disorder. In panels  \textbf{(a)}, \textbf{(b)} and \textbf{(c)} we show a corresponding cut of the local valley Hall marker $\mathfrak{S}_{\alpha}(\bm{r})$, discriminated by sublattice ($\alpha=A,B$), along the direction $\bm{r}=x_c\hat{\mathbf{x}}+y\hat{\mathbf{y}}$ for one particular disorder realization. In \textbf{(a)} and \textbf{(b)} $W=0.2\, t$ and in \textbf{(c)} $\Gamma=0.606$. The insets schematically show the deviations of the flake's on-site energies or the tunneling amplitudes with respect to the pristine case. Panels \textbf{(d)}, \textbf{(e)} and \textbf{(f)} show deviations of the corresponding valley Hall response from the pristine case. The response  (solid blue line) is obtained by averaging along a bulk region of size $L_{\textrm{bulk}}=4.6\,a$ and over 50 different disorder configurations. The standard deviation is indicated in light blue shaded area. Panels \textbf{(g)}, \textbf{(h)} and \textbf{(i)} show the corresponding bulk density of states $\bar \rho_{\textrm{bulk}}$, also averaged over the different disorder realizations, as a function of energy and the disorder intensity $W$ or $\Gamma$. Dashed grey lines indicate the energy of the pLLs in the pristine case. The dashed blue-line shows the chemical potential $\mu_F$ chosen in the upper panels.}
\label{disorder}
\end{figure*}
\textbf{On the effect of disorder, defects and impurities.} In this section, we explore the robustness of the local valley Hall marker in the presence of different types of disorder and lattice imperfections. To this end, we introduce in Eq.~\eqref{H_strain} a general perturbation of the form
\begin{eqnarray}
\hat{H}_\text{dis}^{} &=&\sum_{\bm{r}\in A}V_{\textrm{imp}}(\bm{r})\hat{a}^{\dagger}_{\bm{r}}\hat{a}^{}_{\bm{r}} + \sum_{\bm{r}\in B}V_{\textrm{imp}}(\bm{r})\hat{b}^{\dagger}_{\bm{r}}\hat{b}^{}_{\bm{r}}\\
\notag
& & -\sum_{\bm{r}\in A,j}\delta t_j(\bm{r})\left(\hat{a}^{\dagger}_{\bm{r}}\hat{b}^{}_{\bm{r}+\bm{\delta}_j}+\text{h.c.}\right).
\end{eqnarray}
Here, $V_{\textrm{imp}}(\bm{r})$ represents an on-site potential stemming from the presence of defects or impurities in the lattice. Note that $V_\text{imp}^{}$ can potentially take different values on A and B sites. The term $\delta t_j(\bm{r})$ acts as a bond (off-diagonal) disorder. We will specifically model the latter with a uniform distribution along the entire flake, such that $\delta t_j(\bm{r})/t =  \gamma_j(\bm{r})\tau$, with $\gamma_j(\bm{r})$ chosen randomly in the interval $[-\Gamma,\Gamma]$. This effectively introduces an uncertainty or lack of precision in the space-dependence imprinted on the tunneling amplitudes defined in Eq.~\eqref{hopping_triaxial}.
For the sake of concreteness, two different types of on-site (diagonal) disorder will be separately considered. The first one represents a short-range scattering potential that varies stochastically on the lattice spacing scale, modeled with a number of $N_{\textrm{imp}}$ delta-like impurities chosen out of the $N_{\textrm{tot}}$ total number of sites of the flake, such that 
\begin{equation}
V_{\textrm{imp}}(\bm{r})=V^{\delta}_{\textrm{imp}}(\bm{r})\equiv \sum_{n=1}^{N_{\textrm{imp}}}U_{n}\delta(\bm{r}-\bm{R}_n),  
\end{equation}
with the strengths $U_{n}$ taken randomly within the range $[-W, W]$ and $\bm{R}_n$ the position of the $n$-th impurity. With the aim of considering a scattering potential with longer range, we will also analyze the case of a potential that varies smoothly on the lattice spacing scale. In this last case, the potential profile around each impurity is modeled with a Gaussian function, such that
\begin{equation}
 V_{\textrm{imp}}(\bm{r}) = V^{\lambda}_{\textrm{imp}}(\bm{r})\equiv \sum_{n=1}^{N_{\textrm{imp}}}U_n \textrm{exp}\left(-\frac{|\bm{r}-\bm{R}_n|^2}{2\lambda^2}\right),
\end{equation}
with $\lambda$ characterizing the range of the potential \cite{Marmolejo-Tejada_2018,Adam2009,Klos2010,Ortmann2011}.

In Fig.~\ref{disorder}, we numerically analyze the impact of all these particular lattice imperfections on the valley Hall marker.  
We especially focus on the density response function within the first gap by setting the chemical potential to be $\mu_F = 0.25\,t$ and $\tau=0.07$, as in Figs.~\ref{hexagonal_flakes}\textbf{(a)} and \textbf{(b)}. 
The first and second columns show, respectively, the effect  of the on-site random potential profile $V_{\textrm{imp}}^{\delta}(\bm{r})$ with $N_{\textrm{imp}}=N_{\textrm{tot}}$ and the Gaussian-like profile $V_{\textrm{imp}}^{\lambda}(\bm{r})$, with $N_{\textrm{imp}}=0.01 \,N_{\textrm{tot}}$ and $\lambda = 3\,a$, neglecting the off-diagonal disorder. In the third column, we independently analyze the bond-disordered case. In Figs.~\ref{disorder} \textbf{(a)}, \textbf{(b)} and \textbf{(c)} we show the local valley Hall marker $\mathfrak{S}_{\alpha}(\bm{r})$, discriminated by sublattice ($\alpha=A,B$), along the direction $\bm{r}=x_c\hat{\mathbf{x}}+y\hat{\mathbf{y}}$ for one particular disorder realization in each case scenario. We have chosen disorder strengths that are sufficiently weak compared to the first gap between pLLs, namely defects with maximum on-site energy variations of $W=0.2\,t$ and bond-imperfections with a maximum hopping fluctuation of $0.04\,t$ ($\Gamma=0.606$). The insets schematically show the deviations of the flake's on-site energies or the tunneling amplitudes with respect to the pristine case. Generically, for these moderate values of disorder, the space dependent local valley Hall marker slightly deviates from the one obtained in the perfectly ordered lattice. In Fig.~\ref{disorder}\textbf{(a)}, the fluctuations occur on the lattice spacing scale, while in Fig.~\ref{disorder}\textbf{(b)}, they only take place locally around a Gaussian impurity. Importantly, near the bulk, $\mathfrak{S}_{\alpha}(\bm{r})$ exhibits oscillations around the plateau-like values previously obtained in Fig~\ref{hexagonal_flakes}\textbf{(b)}, which is reminiscent of the results in Ref. \cite{Bianco2011}. We note that when only off-diagonal disorder is present, the bulk response is still polarized in one sublattice ($\mathfrak{S}_{A}(\bm{r})\simeq 0$ around the center of the sample). As opposed to diagonal-disorder, bond-disorder preserves the chiral-symmetry of the Hamiltonian, making the $0^\text{th}_{}$ pLL to remain polarized in one sublattice along with its corresponding density response.

In Figs.~\ref{disorder} \textbf{(d)}, \textbf{(e)} and \textbf{(f)}, we present the corresponding deviations of the average valley Hall response $\overline{\sigma}_V$ from the one obtained in the pristine case $\sigma_V^{0}$. Here, $\overline{\sigma}_V$
is obtained by averaging the local marker in space over $L_{\textrm{bulk}}=4.6\,a$ and over $50$ different disorder realizations.
The light blue shaded area represents the standard error on this average. In each disorder configuration, the intensity of the imperfections is randomly changed and, for the Gaussian-like defects, the position of the impurities in the flake is also aleatory modified. Increasing the disorder intensity naturally enhances the discrepancies with respect to the pristine case in all three scenarios. We note that the standard deviation of the response is visibly less when the impurity concentration is more diluted in the region that is being probed, as shown for a Gaussian potential profile [Fig.~\ref{disorder}\textbf{(b)}].
In Figs.~\ref{disorder}~\textbf{(g)}, \textbf{(h)} and \textbf{(i)}, we show the corresponding bulk density of states averaged over the different disorder realizations ($\bar \rho_\text{bulk}^{}$) as a function of energy and the disorder intensity $W$ or $\Gamma$. Lattice imperfections generically lift the degeneracy of the pLLs, leading to a broadening of these states into bands bearing localized modes around the impurities. An exception takes place for the particular case of off-diagonal disorder, where the energy of the 0-th pLL remains unaltered [see Fig.~\ref{disorder}\textbf{(i)}]. Indeed, one verifies that this chiral-symmetry preserving disorder does not lift the degeneracy of these sublattice polarized states. In all cases, for sufficiently large values of disorder, the mobility gap gets entirely filled with impurity states. In particular, when $W$ is such that the local density of states becomes finite at energy  $\varepsilon = \mu_\text{F}^{} = 0.25\,t$, the corresponding valley Hall coefficient of Figs.~\ref{disorder}\textbf{(d)} and \textbf{(e)} becomes completely erratic.

We remark that a short-range scattering disorder as the one used in Fig.~\ref{disorder}~\textbf{(a)} would be extremely detrimental for a transport experiment relying on the helicity of the valley polarized edge modes, since the presence of delta-like impurities would lead to strong backscattering between the latter~\cite{Low2010}. Conversely, within our scheme, a fairly quantized local valley Hall marker can be obtained as long as the bulk density profiles in the lattice remain sufficiently close to the ones obtained from the analytical pseudo-Landau level states. We also note that, in this proposal, no valley filtering mechanism is required to properly measure a quantized valley Hall coefficient. Indeed, the local density variations with respect to strain have the \textit{same sign} for both valleys [see Eq.~\eqref{Streda_strain}], so both contributions are summed up in the total response.

\section*{Discussion}

This work introduces an alternative approach for measuring the valley Hall response in strained honeycomb lattices, which relies on probing an equilibrium property of these systems locally in the bulk. Specifically, we have demonstrated that quantized valley Hall coefficients can be obtained by measuring the variation of the particle density, deep within the bulk of a sample, upon small variations of the applied strain. This bulk approach to valley Hall physics, which is based on the Widom-St\v{r}eda formula, leads to the introduction of a local valley Hall marker $\mathfrak{S}(\bm{r})$, which is particularly relevant for realistic lattices with open boundary conditions. When properly averaged over a central insulating region, this marker (determined here as a local density response function) remains quantized and robust as a function of both the chemical potential and the pseudo-magnetic field strength. Such a plateau-like behavior takes place whenever the probed region is genuinely incompressible, hence requiring the existence of sufficiently large spectral gaps between pseudo-Landau levels. We have compared our numerical findings with the results expected from a low-energy analytical model that incorporates the effect of strain at lowest order, finding a good agreement for sufficiently large samples and moderate values of strain. In order to obtain strictly quantized (integer) values from this local response, the magnitude of the applied strain should be chosen appropriately:~a  strong strain can potentially lead to large discrepancies with the analytical predictions, while a very weak strain would lead to tiny spectral gaps between the pLLs. As a general rule, a clear separation between the bulk and the boundary states of the sample is needed to obtain a satisfactory quantized result, which occurs whenever the magnetic length remains much smaller than the system size. We have investigated different strain configurations and edge terminations, and we have found that the quantization of our proposed marker remains independent of the edge physics, i.e. independently from the existence of helical edge states living at the boundaries of a finite-size sample. This behavior is rooted in the fact that, within our framework, the valley Hall coefficient is directly extracted from a Fermi sea response. This stands in sharp contrast with usual transport measurements within the linear regime, which only have access to Fermi surface properties~\cite{Marmolejo-Tejada_2018}. We have also analysed how these results are affected by the presence of different lattice imperfections, such as random local impurities and bond-disorder in the finite hexagonal flake. Moderate disorder strengths only slightly modify the bulk density profiles, making the valley Hall marker to remain fairly robust as compared to more fragile transport probes, which strongly rely on preserving the helicity of edge modes~\cite{Low2010}.

Synthetic molecular lattices~\cite{Gomes2012,Polini2013,Swart2017,Drost2017,Khajetoorians2019}, where a two dimensional electron gas is confined to move in a properly designed array of carbon monoxide (CO) molecules, present themselves as an appealing experimental platform where our approach can be tested. This technique has been used to realize honeycomb~\cite{Gomes2012}, Lieb~\cite{Swart2017}, kagome~\cite{Kempkes2019} and Kekule~\cite{Freeney2020} lattices. 
In this type of setups, it is not possible to vary the strain as a tuning knob, but one can create several lattices, each with a different level of strain, and compare the resulting particle densities deep within the bulk of each fabricated lattice. The strength of the pseudo-magnetic field can be easily tuned by changing the position of the CO’s in a honeycomb lattice pattern~\cite{Gomes2012,Polini2013}: by approaching or separating them further away, tunneling can be reduced or enhanced, respectively. Since the position of the CO molecules can be manipulated with atomic precision, a high level of tunability is available. Hence, in this quantum simulator platform, the tunneling amplitude of electrons is tuned solely by the separation between the CO’s, without modifying the lattice parameter~\cite{Khajetoorians2019}. Strained honeycomb lattices (Kekule structures) have already been realized and are within experimental reach~\cite{Freeney2020}. The spectral properties of these systems are usually accessed via STM probes, which yield valuable information on the local density of states. In addition, the STM allows one to probe occupied and unoccupied states, since there can be tunneling of electrons from the sample to the tip or vice-versa. In principle, a tomography of the particle density could be reconstructed by integrating  the LDoS in energy up to the desired Fermi level at each lattice position. In this sense, quantized bulk density responses to strain variations should be experimentally accessible with current technologies. Furthermore, we note that the STM techniques might also resolve the sublattice polarization of the valley Hall response.

A possible alternative is offered by ultracold Fermi gases in optical lattices, where strain can be finely adjusted through well-designed atom-light couplings~\cite{Alba2013,Tian2015,Jamotte2022}, and where the local particle density can be directly measured in-situ~\cite{Greiner2009, Kuhr2010, Greiner2015, Zwierlein2015, Gross2015, Kuhr2015,leonard2022realization}.  Last but not least, recent advances in engineering arbitrary two-dimensional optical tweezer arrays~\cite{Barredo2016,Wang2020,Schymik2020,Ebadi2021, Bakr2022} open yet another route for the study of quantum gases in strained lattices, within a highly controllable and scalable environment. \\

\paragraph*{Contribution statement} M. Jamotte and L. Peralta Gavensky contributed equally to this work. \\

\paragraph*{Acknowledgements}
The authors acknowledge fruitful discussions with Ingmar Swart and thank Étienne Lantagne-Hurtubise for pointing out Ref.~\cite{Lantagne-Hurtubise2020}. We warmly acknowledge Luis E. F. Fo\`a Torres for providing his insightful view on valley Hall signals obtained via multiterminal transport probes and for his comments on the manuscript. Work in Brussels is supported by the ERC Starting Grants TopoCold and LATIS, the Fonds De La Recherche Scientifique (FRS-FNRS, Belgium), the FRIA grant FC 38756 and the EOS grant CHEQS. Work in Padova is supported by the Rita Levi Montalcini Program through the fellowship DI\_L\_LEVI22\_01.

\bibliography{biblio}

\appendix*

\section{Derivation of Eq.~\eqref{Streda_strain_valley}}\label{AppendixA}

Focusing separately on the valley $\xi = \pm 1$, we split the Brillouin zone into two parts: around $\mathbf K'$, $k_y^{} \in [-\pi/\sqrt{3}a,0]$, and  around $\mathbf{K}$, $k_y^{} \in [0,\pi/\sqrt{3}a]$. The contribution of each valley to the bulk particle density is given by
\begin{equation}
    n^\xi_{}(x) = \sum_{p=0}^{\xi M/2} \sum_\varepsilon \theta(\mu_\text{F}^{}-\varepsilon)\,  |\Psi_\varepsilon^{k_y^{}}(x)|^2,
\end{equation} 
where $k_y^{} = 2\pi p/L_y$,  $p \in \{0,...,\xi M/2\}$ and $L_y^{} = \sqrt{3}M a$. Here $\Psi_{\varepsilon}^{k_y}(x)$ is a two-component spinor describing an eigenstate of energy $\varepsilon$ and quasimomentum $k_y$. Each of its components accounts for the weight of the corresponding state on the $\mathcal{A}$ and $\mathcal{B}$ sublattices, respectively.
In the thermodynamic limit,
\begin{equation}\label{key}
\frac{2\pi}{L_y^{}} \sum_{p = 0}^{\xi M/2} \xrightarrow{M^{} \to \infty} \xi \int_{0}^{\xi\pi/\sqrt{3}} dk_y^{},
\end{equation}
and hence
\begin{equation}
\label{A3}
n^\xi_{}(x) = \frac{\xi L_y^{}}{2\pi} \int_{0}^{\xi\pi/\sqrt{3}} dk_y^{} \sum_\varepsilon \theta(\mu_\text{F}^{}-\varepsilon)\,  |\Psi_\varepsilon^{k_y^{}}(x)|^2.
\end{equation} 
For $\mu_\text{F}^{}>0$, $n^\xi_{}(x) = n_-^\xi (x) + n_+^\xi(x) + n_0^\xi(x)$, where $ n_{+}^\xi(x)$ ($ n_{-}^\xi(x)$) is the density of particles in the states with $\varepsilon>0$ ($\varepsilon<0$) and $n_0^\xi(x)$ is the density of particles in the state with zero energy. Note that this can be rewritten as $n^\xi_{}(x) = n_{\textrm{hf}}^\xi (x) + n_+^\xi(x) + n_0^\xi(x)/2$, where $n_{\textrm{hf}}^\xi (x)$ is the density of a half-filled system. When the system is exactly half-filled, the Hall conductivity is strictly zero, meaning that 
\begin{equation}
 \frac{\partial n_\text{hf}^{\xi}}{\partial B^\xi_\tau} = 0.
\end{equation}
Indeed, particle-hole symmetry in this problem ensures that, for this particular filling, the particle density at each cell $n_{\text{hf}}(\bm{r})=1$, meaning that the density variations with respect to strain are exactly zero. The Hall conductivity at each valley can then be obtained as
\begin{equation}
\label{sigma_H_xi}
    \begin{split}
    \sigma_H^{\xi} &= e\frac{\partial n^\xi}{\partial B_\tau^\xi}\Bigg \rvert_{\mu_\text{F}^{}>0}= e\frac{\partial }{\partial B_\tau^\xi} \left(n_+^\xi(x) + \frac{n_0^\xi(x)}{2}\right)\Bigg \rvert_{\mu_\text{F}^{}>0}.
    \end{split}
\end{equation}
Close to each Dirac point $\xi \mathbf K$  (i.e. $|q_y^{}|a\ll 1$) and in the continuum limit $\ell_B^{} \gg a$ (i.e. $\tau \ll 1$), the spinors in Eq.~\eqref{A3} can be approximated with the eigenstates of Eq.~\eqref{H_Dirac_A}, which, in the case of $\tau>0$, are given by
\begin{equation}\label{psi_LL}
\begin{split}
\Psi^{q_y^{}}_\nu (x)= \frac{1}{\sqrt{L_y^{}\ell_B^{}}} \frac{e^{i q_y y}}{\sqrt{2-\delta_{\nu0}^{}}}
\begin{pmatrix}
\phi_{|\nu|}^{}(X)\\
\textrm{sg}(\nu)\phi_{|\nu|-1}^{}(X)
\end{pmatrix},\\
~\\
\end{split}
\end{equation}
where $\phi_\nu^{}(X) = 1/(\sqrt{\pi} 2^\nu \nu!)^{-1/2} e^{-X^2/2} H_\nu(X)$ is the normalized harmonic oscillator wavefunction, $\ell_B^{2} = \hbar/e |B^\xi_\tau|$ and $X\ \equiv (x-x_\text{c}^{})/\ell_B^{}-\xi q_y^{}\ell_B^{}$. The index $\nu$ labels the discrete set of energies of the relativistic pseudo-Landau levels. In order to perform the derivative in Eq.~\eqref{sigma_H_xi} we can explicitly compute
\begin{equation}\label{A6}
\begin{split}
n_{+}^{\xi}(x) + \frac{n_0^\xi(x)}{2} =& \frac{\xi }{4\pi \ell_B^{}} \sum_{\nu = 1}^{\nu_{\max{}}} \int_{-\xi \infty}^{\xi \infty } dq_y^{} \, \\
&\times \left[ |\phi_{\nu}(X)|^2+|\phi_{\nu-1}(X)|^2 \right]\\
&+ \frac{\xi }{4\pi \ell_B^{} } \int_{-\xi \infty}^{\xi \infty } dq_y^{} |\phi_{0}(X)|^2, 
\end{split}
\end{equation}
where $\mu_\text{F}^{}$ lies between the levels $\nu_{\max{}}$ and $\nu_{\max{}}+1$. Note that we  extended the limits of the bounded integral in Eq.~\eqref{A3} to $\pm \xi \infty$ in the above equation. Since the eigenstates in Eq.~\eqref{psi_LL} decay exponentially with $q_y$, this remains a fairly good approximation. By simply using the normalization of the wavefunctions, the integral in Eq.~\eqref{A6} can be readily performed to find that
\begin{equation}\label{key}
n_{+}^\xi (x) + \frac{n_{0}^\xi(x)}{2} \simeq \frac{1}{4\pi \ell_B^2} (2\nu_{\max{}}+1)
= \frac{e  |B^\xi_\tau|}{2\pi \hbar}\left(\nu_{\max{}}+\frac{1}{2}\right). \nonumber
\end{equation}
Eventually, the Hall conductivity at each valley reads
\begin{equation}
    \sigma_H^\xi = e\frac{\partial  n^\xi}{\partial B^\xi_\tau}\Bigg\rvert_{\mu_F} \simeq \xi\left(\nu_{\max{}}^{}+\frac{1}{2}\right) \sigma_0^{}, 
\end{equation}
where $\sigma_0^{} = e^2/h$. The derivation of $\sigma_H^\xi$ is similar for $\mu_\text{F}^{} <0$, where the values of $\nu_{\max{}}$ will be taken negative.

From Eq.~\eqref{A6}, it is clear that each sublattice contributes with a quantized result. Indeed, the summation runs over A and B sites and can can be separated into two sums. The sum over A sites contains the terms $|\phi^{}_\nu(X)|^2$ and $|\phi_0^{}(X)|^2$, while the sum over the B sites only contains the terms $|\phi^{}_{\nu-1}(X)|^2$. Performing the integrals separately and using the wavefunction normalization, one gets
\begin{equation}\label{key}
\begin{split}
n_\text{A}^\xi (x) &= \frac{e  |B^\xi_\tau|}{2\pi \hbar}\left(\frac{\nu_{\max{}}}{2}+\frac{1}{2}\right),\\
n_\text{B}^\xi (x) &= \frac{e  |B^\xi_\tau|}{2\pi \hbar}\frac{\nu_{\max{}}}{2}.
\end{split}
\end{equation}
So,
\begin{equation}
    \begin{split}
        e\frac{\partial  n_\text{A}^{\xi}}{\partial B^\xi_\tau}\Bigg\rvert_{\mu_F} &= \xi \left(\frac{\nu_{\max{}}}{2}+\frac{1}{2} \right) \sigma_0,\\
        e\frac{\partial  n_\text{B}^{\xi}}{\partial B^\xi_\tau}\Bigg\rvert_{\mu_F} &= \xi \frac{\nu_{\max{}}}{2} \sigma_0,
    \end{split}
\end{equation}
which means that
\begin{equation}
    \begin{split}
        \sigma_V^{} &= e\frac{\partial  n_\text{bulk}^{}}{\partial B^K_\tau}\Bigg\rvert_{\mu_F} = \sigma_V^\text{A} + \sigma_V^\text{B},
    \end{split}
\end{equation}
where $\sigma_V^\text{A} = (\nu_{\max{}}^{} +1) \sigma_0$ and $\sigma_V^\text{B} = \nu_{\max{}}^{} \sigma_0$.

\end{document}